\documentclass{aa}  
\usepackage{url}
\usepackage{graphicx}
\usepackage{multirow}
\usepackage{natbib}
\usepackage[breaklinks=true]{hyperref}
\bibpunct{(}{)}{;}{a}{}{,}
\usepackage[varg]{txfonts}
\begin{document}

  \title{Phase-resolved X-ray spectroscopy and spectral energy distribution of
    the X-ray soft polar RS~Caeli
     \thanks{Based on observations obtained with XMM-Newton, an ESA science
     mission with instruments and contributions directly funded by ESA Member
     States and NASA.}
   }



   \author{I.~Traulsen    \inst{\ref{aip}}
     \and  K.~Reinsch     \inst{\ref{iag}}
     \and  A.~D.~Schwope  \inst{\ref{aip}}
     \and  R.~Schwarz     \inst{\ref{aip}}
     \and  F.~M.~Walter   \inst{\ref{suny}}
     \and  V.~Burwitz     \inst{\ref{mpe}}
          }

   \offprints{I.~Traulsen}

   \institute{Leibniz-Institut f\"ur Astrophysik Potsdam (AIP), An der
     Sternwarte 16, 14482 Potsdam, Germany\\
              \email{itraulsen@aip.de}\label{aip}
         \and
              Institut f\"ur Astrophysik, Georg-August-Universit\"at
              G\"ottingen, Friedrich-Hund-Platz 1, 37077 G\"ottingen,
              Germany\label{iag}
         \and
              Department of Physics and Astronomy, Stony Brook University,
              Stony Brook, NY 11794-3800, USA\label{suny}
         \and
              Max-Planck-Institut f\"ur extraterrestrische Physik, P.O.~Box
              1312, 85741 Garching, Germany\label{mpe}
             }

   \date{Received 28 February 2013 / Accepted 6 December 2013}

%
%

  \abstract
   {RS~Cae is the third target in our series of XMM-Newton observations of
     soft X-ray-dominated polars.}
   {Our observational campaign aims to better understand and describe the
     multiwavelength data, the physical properties of the system components,
     and the short- and long-term behavior of the component fluxes in RS~Cae.}
   {We employ stellar atmosphere, stratified accretion-column, and widely used
     X-ray spectral models. We fit the XMM-Newton spectra, model the multiband
     light curves, and opt for a mostly consistent description of the spectral
     energy distribution.}
   {Our XMM-Newton data of RS~Cae are clearly dominated by soft X-ray
     emission. The X-ray light curves are shaped by emission from the main
     accretion region, which is visible over the whole orbital cycle,
     interrupted only by a stream eclipse. The optical light curves are formed
     by cyclotron and stream emission. The XMM-Newton X-ray spectra comprise a
     black-body-like and a plasma component at mean temperatures of 36\,eV and
     7\,keV. The spectral fits give evidence of a partially absorbing and a
     reflection component. Multitemperature models, covering a broader
     temperature range in the X-ray emitting accretion regions, reproduce the
     spectra appropriately well. Including archival data, we describe the
     spectral energy distribution with a combination of models based on a
     consistent set of parameters and derive a lower limit estimate of the
     distance $d\gtrsim 750\,\mathrm{pc}$.}
   {The high bolometric soft-to-hard flux ratios and short-term variability of
     the (X-ray) light curves are characteristic of inhomogeneous
     accretion. RS~Cae clearly belongs in the group of polars that show a very
     strong soft X-ray flux compared to their hard X-ray flux. The different
     black-body fluxes and similar hard X-ray and optical fluxes during the
     XMM-Newton and ROSAT observations show that soft and hard X-ray emission
     are not directly correlated.}

%
%

   \maketitle

%
%
%

%
%
\section{Introduction}

  Thirty X-ray soft polars with negative hardness ratios\footnote{Hardness
    ratio
    $\mathrm{HR1_\mathrm{ROSAT}}=(H_\mathrm{R}-S_\mathrm{R})/(H_\mathrm{R}+S_\mathrm{R})$
    of count rates in the $0.1-0.5\,\mathrm{keV}$ ($S_\mathrm{R}$) and
    $0.5-2.5\,\mathrm{keV}$ ($H_\mathrm{R}$) energy bands, respectively.} HR1
  have been identified in the ROSAT all-sky survey (RASS) source catalog
  \citep{voges:99} by \citet{thomas:98}, \citet{beuermann:99}, and
  \citet{schwope:02}. They are potential members of the subgroup of
  \object{AM~Her}-type systems that show a `soft X-ray excess', i.\,e.,\ a
  significant dominance of soft X-ray ($E \lesssim 0.5\,\mathrm{keV}$) over
  hard X-ray ($E \gtrsim 0.5\,\mathrm{keV}$) luminosity, reviewed for example
  by \citet{ramsay:94}, \citet{beuermann:94}, and
  \citet{beuermann:95}. \citet{ramsay:04ebalance} demonstrated the dependence
  of the soft-to-hard luminosity ratios on calibration, geometrical effects,
  and spectral models. Using accretion-column models \citep{cropper:99} for
  recalibrated ROSAT and XMM-Newton spectra, they came to the conclusion that
  fewer systems show a distinct soft X-ray excess than originally estimated
  from the ROSAT detections.


  \object{RS~Caeli} was among the softest X-ray sources at the epoch of the
  RASS observations, with a hardness ratio of $\mathrm{HR1}=-1.00(1)$. It was
  detected as an extreme ultraviolet source in the ROSAT/WFC and in the EUVE
  surveys \citep{pounds:93,bowyer:94}. \citet{burwitz:96} published the first
  pointed X-ray and additional optical observations of RS~Cae. They derived an
  optical apparent magnitude of $m_V\sim19^\mathrm{m}$, an X-ray flux on the
  order of $10^{-11}\,\mathrm{erg\,cm^{-2}\,s^{-1}}$ in the ROSAT energy band,
  a distance to the binary of at least 440\,pc, and a magnetic field strength
  $B=36(1)\,\mathrm{MG}$ for the white-dwarf primary. The pronounced
  phase-dependent cyclotron harmonics in the optical spectra could be modeled
  for two possible accretion geometries: a binary inclination of $i\sim
  60\degr$ and a colatitude of the accretion region of $\beta\sim 25\degr$, or
  $i\sim 25\degr$ and $\beta\sim60\degr$. In the first case, stream absorption
  dips should be seen in the X-ray light curves. They found two candidate
  orbital periods of $0\fd0708(14)$ or $0\fd0652(15)$, giving preference to
  the longer one. The optical spectra showed no clear signature of the M-star
  secondary.

  By exploiting new XMM-Newton data of polars that had not been observed in
  X-rays since ROSAT, we are studying their system properties and the energy
  balance, in particular during high states (\object{AI~Tri},
  \citealt{traulsen:10}, RS~Cae, this work) and intermediate high states of
  accretion (AI~Tri, \object{QS~Tel}, \citealt{traulsen:10,traulsen:11}).  In
  this paper, we present our third pointed XMM-Newton observation of a soft
  X-ray selected polar. Section~\ref{sec:data} introduces the X-ray and
  optical data of RS~Cae on which our analysis is based. In
  Sect.~\ref{sec:photo}, we describe the multiwavelength light curves and
  confirm the orbital binary period of $0\fd071$. Section~\ref{sec:spectra} is
  dedicated to the X-ray spectra, the spectral models, and the derived
  parameters. The whole spectral energy distribution, an approach to a
  consistent modeling of spectra and light curves, and the implications on the
  system geometry are presented in Sect.~\ref{sec:disc_sedmodeling}. We close
  the paper with a discussion of the component fluxes and the energy budget of
  RS~Cae in Sect.~\ref{sec:disc_energy}.

  \begin{table}
    \caption{Barycentric timings and $1\sigma$ errors of the dip centers in
      the XMM-Newton light curves of RS~Cae.}
    \label{tab:minima}\centering
    \begin{tabular}{c@{\qquad}c@{\qquad}r}
      \hline\hline
      $\mathrm{BJD}_\mathrm{min}$(TT) & $\Delta T_\mathrm{min}$
      & $O-C~$ \\ \hline
     $2\,454\,903.05846$  &  $0.00044$  & $-0.0123$ \\
     $2\,454\,903.12995$  &  $0.00031$  & $-0.0042$ \\
     $2\,454\,903.20209$  &  $0.00037$  & $ 0.0130$ \\
     $2\,454\,903.27310$  &  $0.00031$  & $ 0.0143$ \\
     $2\,454\,903.34195$  &  $0.00043$  & $-0.0148$ \\
     $2\,454\,903.41275$  &  $0.00056$  & $-0.0165$ \\
      \hline
    \end{tabular}
  \end{table}

%
%
\section{Observations and data reduction}
\label{sec:data}

  RS~Cae was scheduled for a 50\,ks observation with XMM-Newton on March
  12/13, 2009 (observation ID 0554740801). Owing to background radiation, the
  EPIC/pn and MOS exposures had to be stopped after 35\,ks and 39\,ks,
  respectively. All three EPIC detectors were operated in full frame mode with
  the thin filter. Using standard \textsc{sas}\,v9.0 tasks, we extracted light
  curves and spectra from circular source regions on the EPIC chips with a
  radius of 27.5\,arcsec for EPIC/pn and of 22.5\,arcsec for EPIC/MOS. We used
  large circular background regions with radii between 75 and 100\,arcsec on
  the same chip as the source for the background correction. Spectra were
  taken from the first 28\,ks of the exposure, excluding the high-background
  intervals. During our pointing, the source reached net peak count
  rates\footnote{maximum rate and Poissonian error derived from 1\,ks
    light-curve segments} of $1.64\pm0.04\,\mathrm{cts\,s^{-1}}$ for EPIC/pn,
  $0.16\pm0.01\,\mathrm{cts\,s^{-1}}$ for MOS1, and
  $0.24\pm0.02\,\mathrm{cts\,s^{-1}}$ for MOS2, which is in the range that can
  be expected from the ROSAT All-Sky Survey results \citep{voges:99} for a
  high-state observation. The net source count rate measured with RGS was
  consistent with zero.

  With the optical monitor OM, we performed fast-mode photometry consecutively
  in the 3000$-$3900\,{\AA} band using the $U$ filter (mean net count
  rate\footnote{corrected count rates and errors given by the OM
    source-detection tasks} $1.06\pm0.02\,\mathrm{cts\,s^{-1}}$), in the
  2450$-$3200\,{\AA} band using the UVW1
  ($0.53\pm0.02\,\mathrm{cts\,s^{-1}}$), and in the 2050$-$2450\,{\AA} band
  using the UVM2 filter ($0.13\pm0.01\,\mathrm{cts\,s^{-1}}$). The exposure
  times of 8.2\,ks per light-curve segment corresponded to about 1.3 orbital
  cycles. We extracted fast-mode light curves using the \textsc{sas} v10.0
  version of the source detection algorithm and the task \textsc{omfchain} and
  took the background information from the imaging data. The UVM2 data were
  mostly affected by the increased background. For the rest of the visit, the
  optical monitor was operated with the grism1 filter
  (2000$-$3500\,{\AA}). This part of the observation fell completely in the
  time interval of high background radiation. Three of the fourteen scheduled
  800\,s exposures were taken, but are unusable due to a low signal-to-noise
  ratio.

  To trigger the XMM-Newton observation, we repeatedly obtained optical
  photometry of RS~Cae between December 2008 and March 2009 at the CTIO 1.3\,m
  telescope/ANDICAM, operated by the SMARTS consortium. Two of the $B$-band
  light curves cover more than one orbital period: one taken on September 22,
  2008 during a low state of accretion and one taken simultaneously to the
  last part of the XMM-Newton observations. Each of them comprises 32 data
  points with integration times of 180\,s and a time resolution of
  226$-$227\,s. Photometry was done relative to stars in the field as
  described in \citet{gerke:06}.

  We converted the ground-based data from UTC to terrestrial time TT,
  consistently with the satellite data, and corrected all times used in this
  paper to the barycenter of the solar system using the JPL ephemeris
  \citep{standish:98}.

%
%
\section{Multiband light curves and photometric period}
\label{sec:photo}

\begin{figure}
    \centering
    \includegraphics[width=8.8cm]{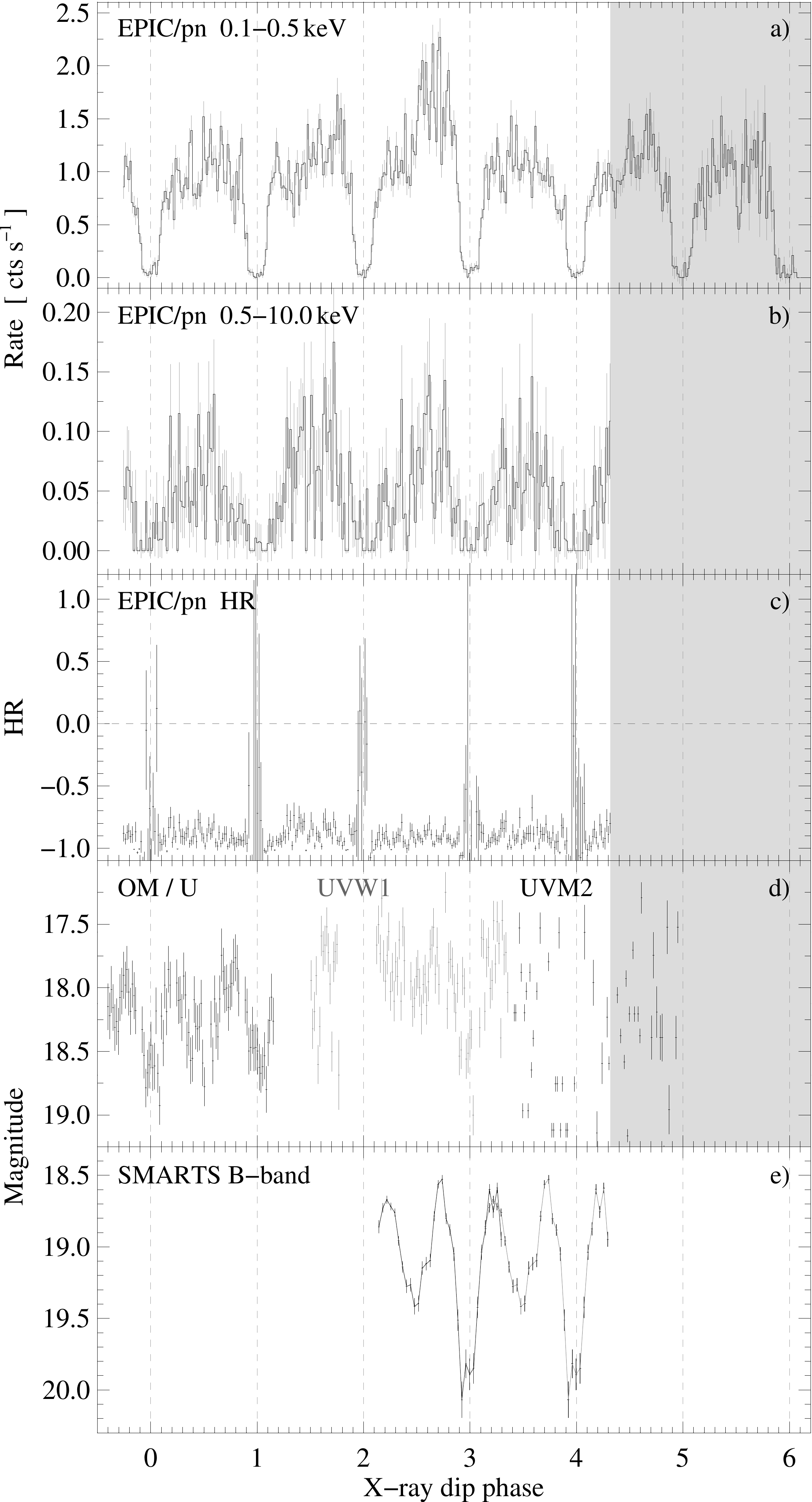}
    \caption{September 2009 light curves of RS~Cae in time bins of 100\,s,
      folded on the orbital period with the center of the X-ray dips defining
      phase zero (Eq.~\ref{eq:ephem}). The grey area marks the interval of
      high background activity during the XMM-Newton pointing. \textbf{a--b)}
      Energy-resolved EPIC/pn light curves. \textbf{c)} Corresponding hardness
      ratios
      $\mathrm{HR}_\mathrm{XMM}=(H_\mathrm{X}-S_\mathrm{X})/(H_\mathrm{X}+S_\mathrm{X})$.
      \textbf{d)} Optical and ultraviolet light curves measured subsequently
      with three filters at the optical monitor. \textbf{e)} Optical $B$-band
      light curve with a time resolution of about 227\,s, plotted twice and
      shifted by $-3$ and $-2$ orbital cycles, respectively.}
    \label{fig:multilc}%
  \end{figure}

\begin{figure}
    \centering
    \includegraphics[width=7.7cm]{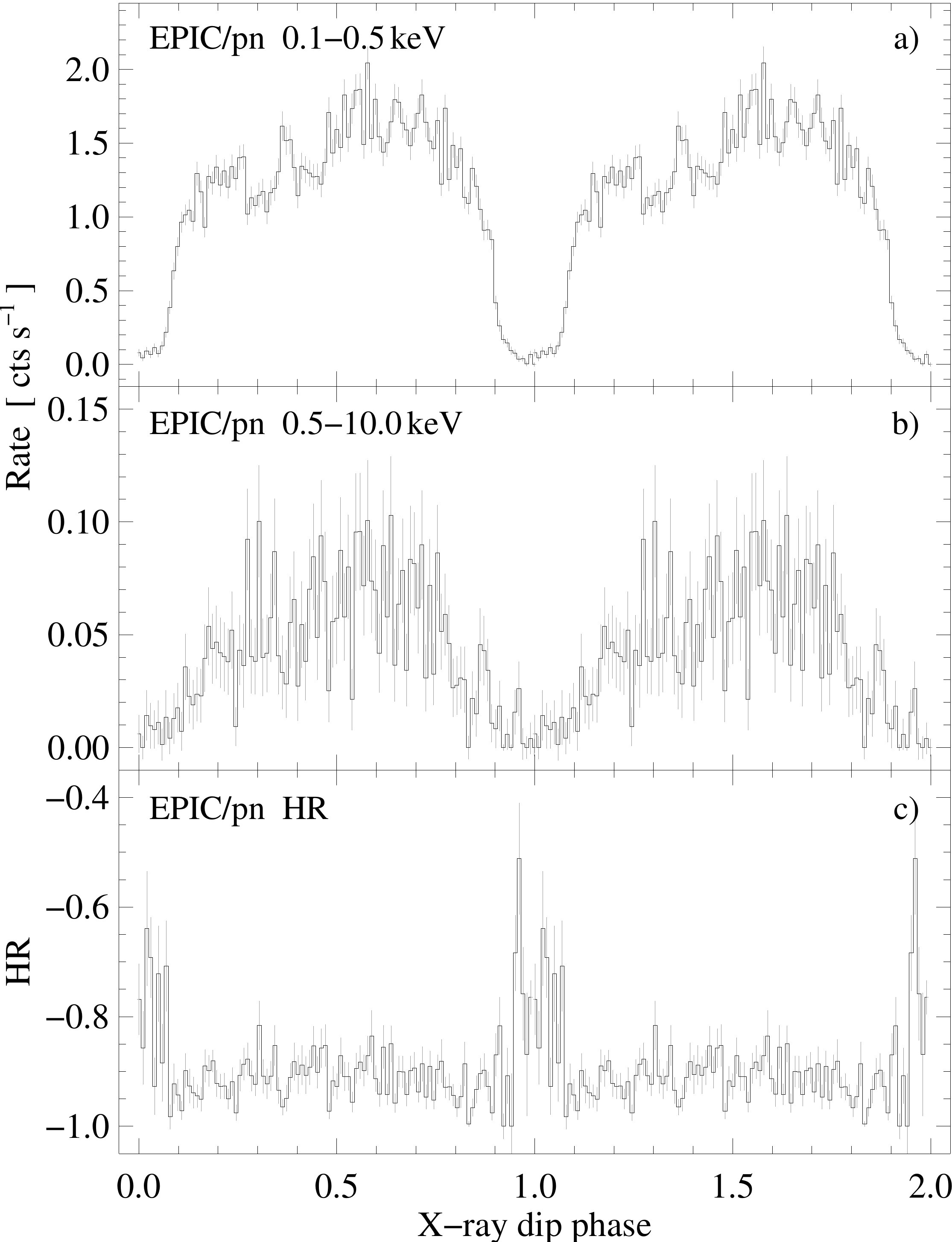}
    \caption{Phase-averaged X-ray light curves, excluding the high-background
      interval, in time bins of 60\,s and with the same phase convention and
      energy ranges as in Fig.~\ref{fig:multilc}. \textbf{a--b)}
      Energy-resolved EPIC/pn light curves. \textbf{c)} Corresponding hardness
      ratios.}
    \label{fig:lcprofile}%
  \end{figure}

  Figure~\ref{fig:multilc} gives a synopsis of the X-ray, ultraviolet, and
  optical light curves of RS~Cae. The corresponding X-ray light-curve profiles
  are shown in Fig.~\ref{fig:lcprofile}. Our phase convention refers to the
  centers of the pronounced X-ray dips and is derived in
  Sect.~\ref{sec:ephem}.

\begin{figure}
    \includegraphics[width=8.3cm]{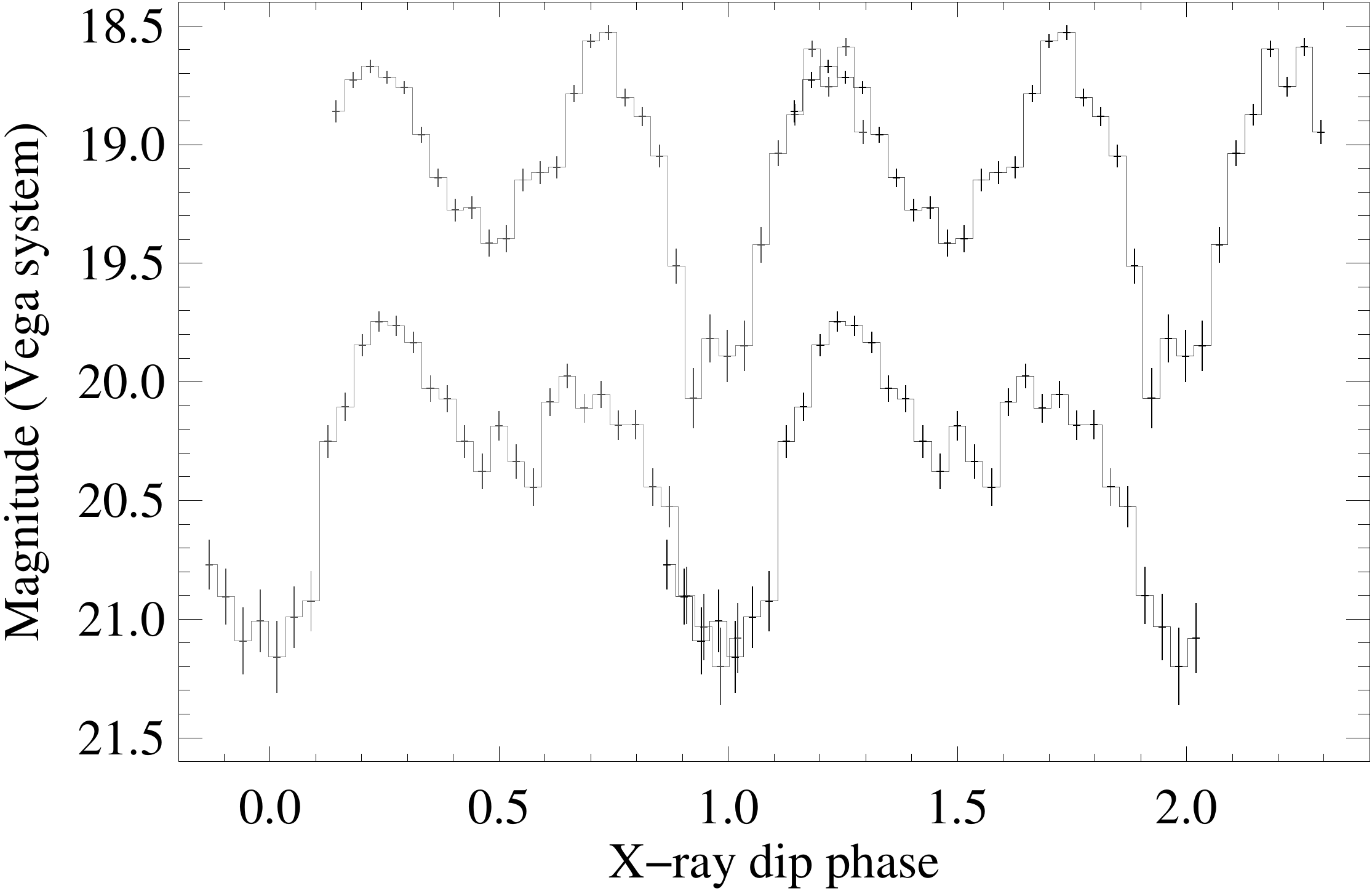}
    \caption{SMARTS $B$-band light curves of RS~Cae obtained at 2008/09/22
      during a low state and at 2009/03/12 during a high state of accretion,
      each folded on a period of $0\fd07089$ and plotted twice.}
    \label{fig:optlcs}%
  \end{figure}

  The XMM-Newton X-ray light curves (Figs.~\ref{fig:multilc}a, b, and
  \ref{fig:lcprofile}) show clear periodicity that could not be detected in
  the 1992 ROSAT / PSPC light-curve segments presented by \citet{burwitz:96}.
  With EPIC/pn, about 21\,400 source counts were collected in the soft X-ray
  band at energies below 0.5\,keV and about 760 in the hard X-ray band at
  energies above 0.5\,keV. Correspondingly, the hardness ratios
  $\mathrm{HR}_\mathrm{XMM}=(H_\mathrm{X}-S_\mathrm{X})/(H_\mathrm{X}+S_\mathrm{X})$
  between hard ($H_\mathrm{X}, 0.5-10.0\,\mathrm{keV}$) and soft
  ($S_\mathrm{X}, 0.1-0.5\,\mathrm{keV}$) counts are remarkably low
  (Fig.~\ref{fig:multilc}c). During a sharp recurring light-curve dip, the
  EPIC count rates almost drop to zero for 0.07 in phase, and the hardness
  ratios increase. This dip is visible in all X-ray bands and coincides with
  ultraviolet and optical light-curve minima.  The optical and ultraviolet
  light curves (Figs.~\ref{fig:multilc}d and e) are double-humped with
  semi-amplitudes between 0.5 and 0.7\,mag,showing similar shapes during high
  and low states of accretion (Fig.~\ref{fig:optlcs}).  A minimum with a depth
  of about $\Delta\mathrm{mag}\sim 0.8$ occurs at the time of maximum X-ray
  flux. They resemble the double-humped white light curves of
  \citet{burwitz:96}, taken in October 1992 and September 1993.

%
\subsection{X-ray ephemeris}
\label{sec:ephem}

  \citet{burwitz:96} determined a spectroscopic ephemeris for the system and
  referred to the times of minimum spectral flux as phase zero. Using the
  recurrent X-ray dips to constrain the binary period, we can confirm their
  preferred period of $P_\mathrm{orb} = 102\,\mathrm{min} = 0.071\,\mathrm{d}$
  independently by Lomb-Scargle analysis, epoch folding, and a least-squares
  method to compute the maximum of $1/\Sigma(O-C)^2$ (observed minus
  calculated dip times). Observed dip times and $1\sigma$ errors are derived
  by Gaussian fits to $\Delta\varphi\sim 0.2$ segments in the 10s-binned
  EPIC/pn light curve (Table~\ref{tab:minima}). The photometric ephemeris is
  calculated by an error-weighted fit to the observed dip times using
  1ms-spaced trial periods within a $\pm3\sigma$ search interval around the
  longer period of \citet{burwitz:96}.
  \begin{equation}
    \label{eq:ephem}
    \mathrm{BJD}_\mathrm{dip}(\mathrm{TT}) = 2\,454\,903\fd2012(4) +
    0\fd0709(3) \times E
  \end{equation}
  defines phase zero throughout the paper. The uncertainties of the last
  digit, given in parentheses, are $1\sigma$ errors of the $O-C$ method.

%
\subsection{The nature of the soft X-ray dip}
\label{sec:Xraydip}

\begin{figure}
    \centering
    \includegraphics[width=8.8cm]{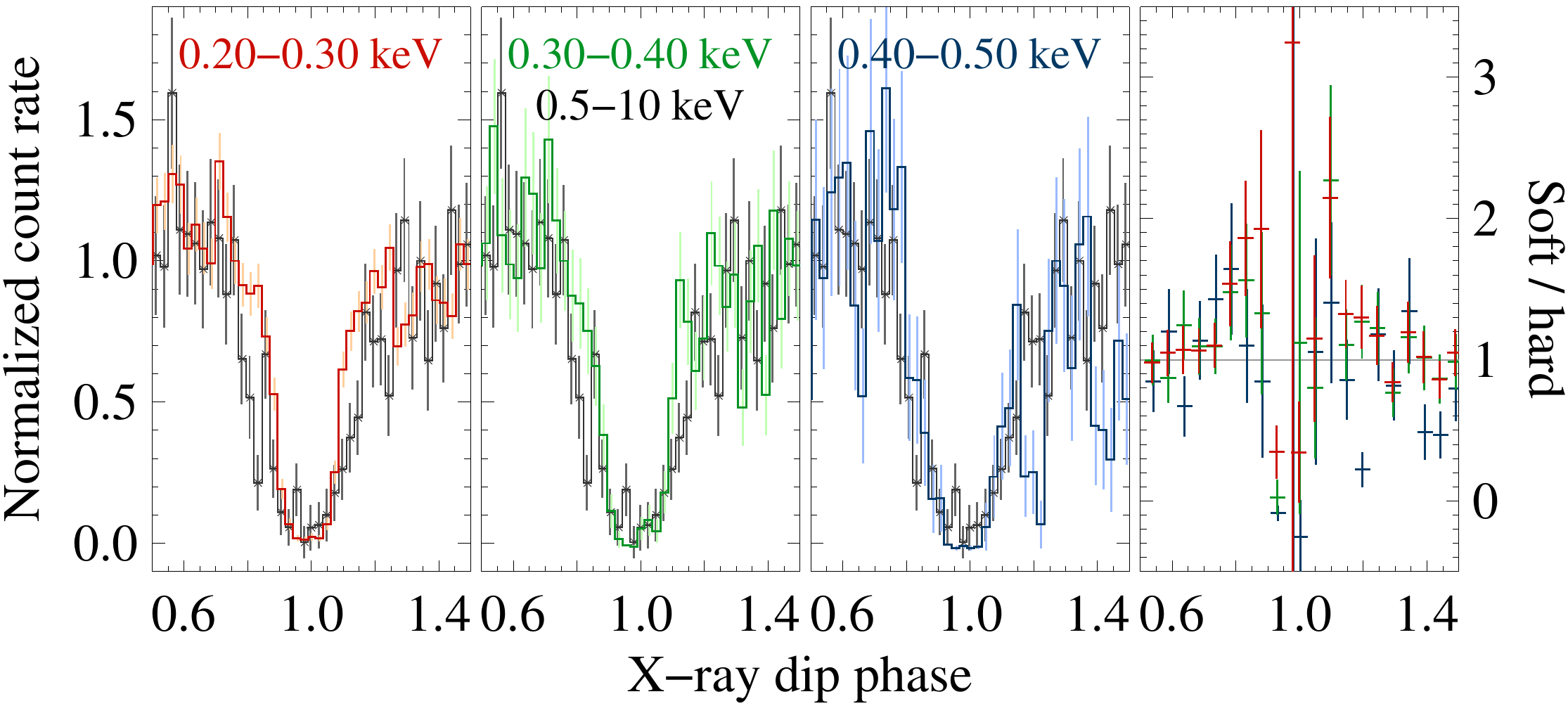}
    \caption{The dip in phase-averaged, energy-resolved X-ray light
      curves. Left to right: Light curves extracted from soft X-ray energy
      intervals of constant width, compared to the harder
      $0.5-10.0\,\mathrm{keV}$ band. The data are binned into a time
      resolution of 150\,s and normalized to their nondip median count
      rates. Right panel: Corresponding ratios of X-ray soft to X-ray hard
      light curves, binned into a time resolution of 300\,s.}
    \label{fig:lcdips}%
  \end{figure}

  Light-curve dips may occur due to an eclipse by the secondary star, a
  self-eclipse of the accretion region, or stream absorption. A self-eclipse
  is unlikely for a geometry with $i+\beta<90\degr$, which is expected for
  RS~Cae. The dips in the optical and UV light curves around phase zero
  resemble a (partial) eclipse feature. In Sect.~\ref{sec:disc_lcsim},
  however, we show that the modulation of these light curves is mostly due to
  cyclotron emission and that their minima can be explained without assuming
  an eclipse by the secondary. We conclude that the sharp dip in the X-ray
  light curves is caused by absorption in the accretion stream when it crosses
  our line of sight towards the accretion region: \textit{(i)} The hardness
  ratios increase rapidly at dip times, where photons at lower X-ray energies
  are more strongly absorbed than those at higher energies. \textit{(ii)} The
  depth of the minima varies slightly from cycle to cycle. \textit{(iii)}
  Accretion-stream and absorption models for RS~Cae following
  \citet{silva:12pre} reproduce the X-ray dip. The X-ray dip appears to be
  broader toward higher energies (Fig.~\ref{fig:lcdips}). The energy
  dependence may indicate that more extended, cooler parts of the accretion
  region are still visible, while the harder emission region is absorbed by
  the dense core of the stream.

%

\section{X-ray spectroscopy}
\label{sec:spectra}

  From the first 28\,ks of our XMM-Newton exposure, we extract EPIC/pn, MOS1,
  and MOS2 spectra as described in Sect.~\ref{sec:data}, excluding the phases
  around the X-ray light curve dips $\varphi_\mathrm{X-ray}\sim 0.9-1.1$, and
  fit them simultaneously to derive general system parameters such as
  temperatures in the accretion region, plasma abundances, and the amount of
  intrinsic absorption.  The EPIC spectra show the typical components of
  ROSAT-discovered polars: \textit{(i)} the black-body like, X-ray soft
  emission, which can be attributed to the accretion-heated surface of the
  white dwarf (Sect.~\ref{sec:softspec}), and \textit{(ii)} the
  bremsstrahlung-like, X-ray hard component, which can be attributed to the
  hot accretion column above the white-dwarf surface
  (Sect.~\ref{sec:hardspec}). Since a very strong soft X-ray component was
  seen in the ROSAT/PSPC spectra and almost $97\,\%$ of the EPIC/pn source
  counts are measured at energies below 0.5\,keV, we are investigating in
  particular the soft and hard X-ray fluxes that we derive from the different
  spectral model approaches and discuss in Sect.~\ref{sec:disc_energy}.

  Our models in \textsc{xspec} v12.6 \citep{arnaud:96,dorman:03} consist of
  two additive spectral components and up to two absorption terms:
  \textit{(i)} the black-body-like component to describe the soft part of the
  spectrum, dominating at energies up to about 0.5\,keV; and \textit{(ii)} the
  plasma component to describe the hard part of the spectrum, dominating from
  energies around $0.5-0.7$\,keV onward. Plasma abundances are given with
  respect to the solar abundances of \citet{asplund:09}.  In addition,
  absorption by material on our line of sight is expected to affect the
  emitted spectra: \textit{(i)} absorption by the interstellar medium. We fit
  it with a \textsc{tbnew}\footnote{Most recent version of \textsc{tbvarabs}
    in
    \textsc{xspec}. See\\ \href{http://pulsar.sternwarte.uni-erlangen.de/wilms/research/tbabs/}{http://pulsar.sternwarte.uni-erlangen.de/wilms/research/tbabs/}.}
  component and employ the abundances of \citet{wilms:00} and cross-sections
  of \citet{verner:96a} and \citet{verner:96b} for it; \textit{(ii)}
  absorption by diffuse gas around the X-ray emitting accretion regions, whose
  amount can vary over the orbital cycle. We fit it with the partially
  covering absorption model \textsc{pcfabs}. Reflection from the white-dwarf
  surface may also contribute to the flux at higher energies. We consider it
  optionally in the fits in Sect.~\ref{sec:hardspec}.

\subsection{The black-body-like component}
\label{sec:softspec}

\begin{figure}
    \centering
    \includegraphics[width=8.8cm]{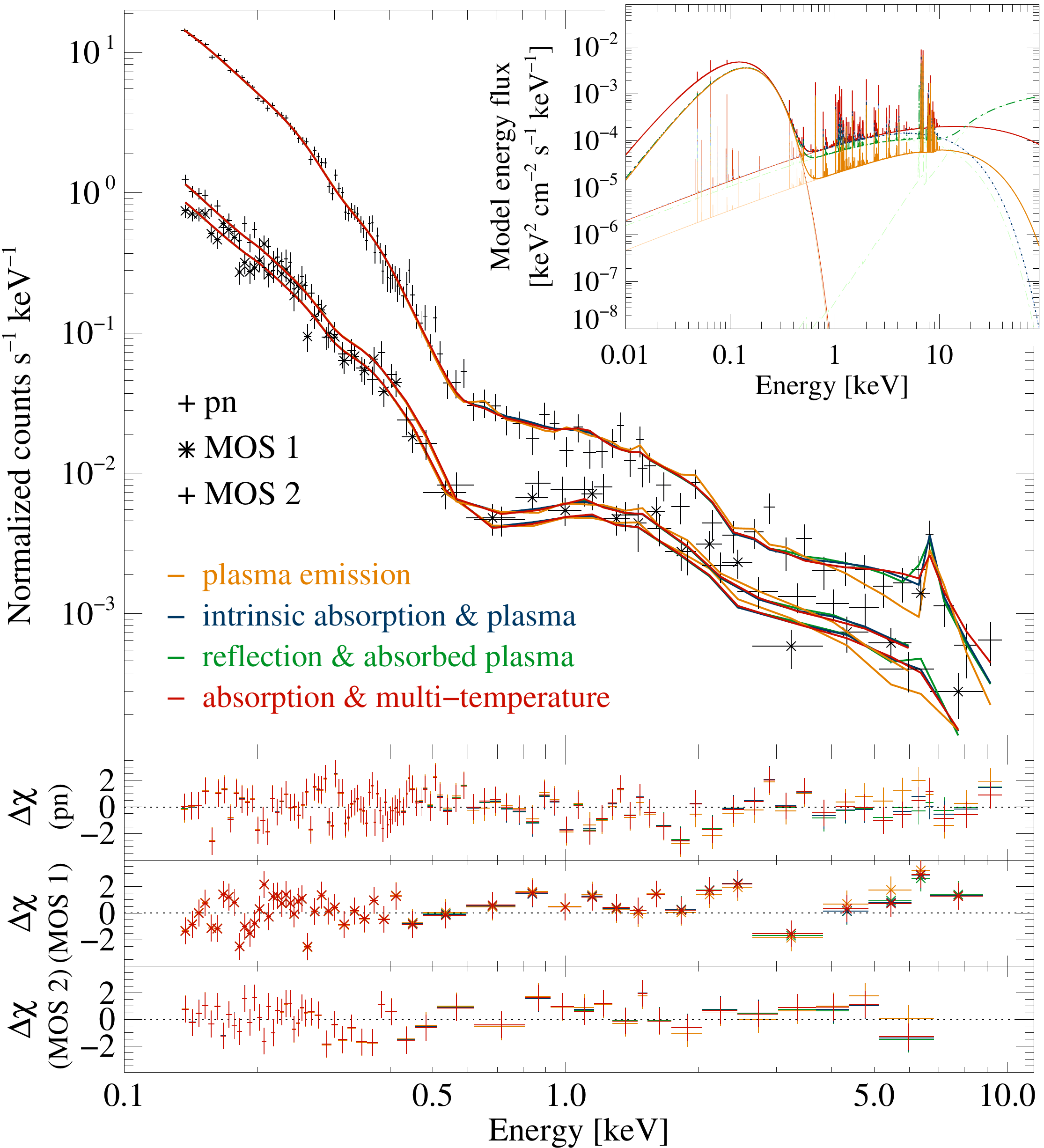}
    \caption{Nondip EPIC spectra of RS~Cae and the \textsc{xspec} black-body
      plus plasma fits. \textbf{Small panel:} Corresponding unabsorbed model
      fluxes.}
    \label{fig:spectra}%
  \end{figure}

  The soft X-ray component is fitted by an absorbed single-temperature black
  body at a temperature of $35.7^{+0.6}_{-0.7}\,\mathrm{eV}$. The
  \textsc{tbnew} absorption term of
  $N_\textsc{H,tbnew}=2.1^{+1.0}_{-0.7}\times 10^{19}\,\mathrm{cm}^{-2}$ stays
  below the upper limit of interstellar hydrogen absorption on our line of
  sight, $N_\textsc{H}=1.2-1.6\times 10^{20}\,\mathrm{cm}^{-2}$
  \citep{kalberla:05,dickey:90}. Using a \textsc{mekal} component for the
  harder part of the spectrum (cf.\ Sect.~\ref{sec:hardspec}), we achieve a
  reduced $\chi^2_\mathrm{red}$ of 1.23 at 199 degrees of freedom over the
  whole XMM-Newton energy range. Residuals remain around the energies of the
  emission lines of helium-like C\,V, N\,VI, and O\,VII
  (Fig.~\ref{fig:spectra}).

  Single-temperature models approximate a time and spatially averaged
  temperature of a rather complex accretion area, which is expected to be
  extended, comprising a wider spread of temperatures, and not necessarily
  circular \citep[cf.][]{milgrom:75,kuijpers:82,ferrario:90}.  Aiming at a
  better description of the temperature gradient in the accretion region on
  the white dwarf, we tested multitemperature black-body models. A fit with a
  second black body shows that the resolution of the data allows for the
  application of multicomponent models. It results in a reduced
  $\chi^2_\mathrm{red}$ of 1.22 (197 d.\,o.\,f., \textsc{mekal} plasma
  component) and an \textit{F-test} probability of $75\,\%$ that the fit is
  improved.
  In the fits with multiple black bodies, we employed models that have been
  successfully used for other polars: black-body components whose effective
  emitting surface areas obey an exponential distribution over temperatures
  \citep[AM~Her,][]{beuermann:12} or whose individual temperatures obey a
  Gaussian distribution over emitting radii
  \citep[AI~Tri,][]{traulsen:10}. The model with a Gaussian temperature
  distribution reproduces the low-energy continuum better, without changing
  the reduced $\chi^2_\mathrm{red}=1.22$ over the full
  $0.1-10.0\,\mathrm{keV}$ range.

  To test the quality of the fits independently of the $\chi^2$ fit
  statistics, we performed a \textit{runs test} for randomness for each model
  and determined the probability that the residuals of the fit are randomly
  distributed around zero. It increases from $P_\mathrm{random}=35\,\%$ for
  the single-temperature black body to $45\,\%$ for the two black-body
  components and to $76\,\%$ for the Gaussian temperature distribution, mainly
  triggered by a low number of residuals changing sign and indicating a
  somewhat higher preference for the multitemperature
  approach.\footnote{Probabilities of the two-tailed \textit{runs test} have
    been calculated as twice the single-tail values of a normalized Normal
    distribution. If the data are described well by the model, we expect a
    random distribution, while systematic deviations manifest themselves as
    longer sequences of positive or of negative residuals and a lower
    probability value.} It yields temperatures up to
  $kT_{\textsc{bbody,}\mathrm{max}}=39.1^{+0.4}_{-0.9}\,\mathrm{eV}$ with
  highest flux at 34.5\,eV and an interstellar absorption term of
  $N_\textsc{H,tbnew}=3.3^{+1.1}_{-0.8}\times 10^{19}\,\mathrm{cm}^{-2}$.
  Owing to the wider temperature range covered, the bolometric model flux
  $1.2^{+0.2}_{-0.1}\times 10^{-11}\,\mathrm{erg\,cm}^{-2}\,\mathrm{s}^{-1}$
  is about $50\,\%$ higher than for the single temperature.

\begin{table*}
    \caption{Best-fit parameters for the non-dip EPIC spectra of RS~Cae,
      employing single-temperature components
      \textsc{tbnew(bbody+\,}\textit{plasma}\textsc{)}.}
    \label{tab:xspecfits} 
    \centering
    \begin{tabular}{l*{3}{c@{\quad}}r@{}l@{\quad}c@{\quad}r@{.}lcrc}
      \hline\hline\\[-2ex]
      Plasma component & $\chi^2_\mathrm{red}$ & $N_\textsc{H,tbnew}$ &
      $kT_\textsc{bbody}$ & \multicolumn{2}{c}{$N_\textsc{H,pcfabs}$} &
      cover. & \multicolumn{2}{c}{$kT_{\textsc{mekal}}$} & abund.  &
      $F_\mathrm{bol}(\textsc{bbody})$ & $F_\mathrm{bol}(\textsc{mekal})$ \\
      & & [$10^{19}\,\mathrm{cm}^{-2}$] & [$\mathrm{eV}$] &
      \multicolumn{2}{@{}c@{\quad}}{[$10^{23}\,\mathrm{cm}^{-2}$]} & [$\%$] &
      \multicolumn{2}{c}{[$\mathrm{keV}$]} & (solar) &
      \multicolumn{2}{c}{[$10^{-12}\,\mathrm{erg\,cm}^{-2}\,\mathrm{s}^{-1}$]}
      \\

      \hline\\[-1.5ex]

      \textsc{mekal}
      & 1.30 & $2.1^{+0.8}_{-0.7}$ & $35.6^{+0.6}_{-0.7}$ & & & &
      $12$&$6^{+10.6}_{-2.7}$ & $7.8^{+5.4}_{-2.7}$ &
      $7.7^{+0.8}_{-0.4}$ & $0.32^{+0.09}_{-0.05}$ 
      \\[.8ex]

      \textsc{pcfabs(mekal)}
      & 1.23 & $2.1^{+1.0}_{-0.7}$ & $35.7^{+0.6}_{-0.7}$ &
      \quad$3$&$.9^{+6.0}_{-2.0}$ & $72^{+8}_{-12}$ & $7$&$4^{+3.6}_{-2.6}$ &
      $1.0^{+1.6}_{-0.7}$ & $7.9^{+1.0}_{-0.8}$ & $0.73^{+0.05}_{-0.05}$
      \\[.8ex]

      \textsc{pexmon+pcfabs(mekal)}
      & 1.23 & $2.2^{+0.9}_{-0.7}$ & $35.7^{+0.7}_{-0.7}$ & $2$&$.9^{+11.9}_{-1.9}$ 
      & $62^{+26}_{-20}$ & $6$&$8^{+3.3}_{-2.1}$ & $:= 1.0$ &
      $7.9^{+0.5}_{-0.5}$ & $0.52^{+0.04}_{-0.04}$
      \\[.3ex] \hline
    \end{tabular}
    \tablefoot{%
      \textsc{pexmon} component: inclination $i:=65^\circ$, photon index
      $1.1$, scaling factor $-0.7$. Unabsorbed bolometric fluxes have been
      determined via \textsc{cflux} within \textsc{xspec}. Errors are given
      within a 90\,\% confidence range.%
      }


\end{table*}

\subsection{The plasma component} 
\label{sec:hardspec}

  The hard X-ray component is fitted by a \textsc{mekal} plasma model
  \citep[cf.][]{mewe:85,liedahl:95}, comprising continuum and line
  emission. Table~\ref{tab:xspecfits} summarizes the parameters of our three
  main \textsc{mekal} fits, which we now describe. The pure \textsc{mekal}
  model is not sufficient to reproduce the observed iron emission around
  6.7\,keV. Underestimating the continuum flux, it results in an
  unrealistically high element abundance of several times the solar values. We
  include a partially covering absorber \textsc{pcfabs} as the simplest
  approach to the complex absorption spectrum expected for the accretion
  emission \citep[cf.][]{done:98,cropper:98}. The best fit yields a mean
  plasma temperature of $kT_\textsc{mekal}=7.4^{+3.6}_{-2.6}\,\mathrm{keV}$ at
  solar element abundances and intrinsic absorption of
  $N_\textsc{H,pcfabs}=3.9^{+6.0}_{-2.0}\times 10^{23}\,\mathrm{cm}^{-2}$,
  covering $72^{+8}_{-12}\,\%$ of the emission region. The bolometric model
  flux of the \textsc{mekal} component increases by a factor of about three
  when adding the absorption term, as obvious from the plot of the unabsorbed
  model components in Fig.~\ref{fig:spectra}.

  In addition, we test the spectra for a neutral Compton reflection
  component. Reflection features were detected, for example, by
  \citet{beardmore:95} for AM~Her, by \citet{done:98} for \object{BY~Cam}, or
  by \citet{bernardini:12} for hard X-ray selected polars, and theoretically
  investigated by \citet{teeseling:96,matt:04,mcnamara:08}. In the EPIC
  spectra of RS~Cae, there is no direct evidence of a Fe K$\alpha$ fluorescent
  line at 6.4\,keV, since the components of the line complex are resolved.  We
  find an upper limit on the order of $2\times
  10^{-5}\,\mathrm{photons\,cm}^{-2}\,\mathrm{s}^{-1}$ in a Gaussian emission
  line component at 6.4\,keV. To test for a reflection continuum, we employed
  the model by \citet{nandra:07}, developed for Compton reflection by neutral
  gas in AGN with a power law as incident spectrum, which is based on a model
  by \citet{magdziarz:95} and implemented as \textsc{pexmon} in
  \textsc{xspec}. A higher \textit{runs-test} probability of
  $P_\textrm{random}=56\,\%$, compared to $38\,\%$ for
  \textsc{bbody+pcfabs(mekal)}, indicates that a reflection component might be
  present.
 


  As described for the accretion region on the white dwarf, a physically
  realistic model for the post-shock accretion column needs to include its
  temperature, density, and velocity structure
  \citep[cf.][]{cropper:99,fischer:01}.  We are employing multitemperature
  accretion-column models that are based on the models of \citet{fischer:01}
  and described by \citet{traulsen:10} for AI~Tri. They fit the spectra of
  RS~Cae well, without improving the $\chi^2$ statistical values of the fit
  significantly. We refer to them in the discussion of the energy balance in
  Sect.~\ref{sec:disc_energy}.  Models with a magnetic field strength of
  $B=36\,\mathrm{MG}$ \citep{burwitz:96} and different specific mass flow
  rates $\dot{m}$ result in similar values of reduced
  $\chi^2_\mathrm{red}=1.2$ to the single-temperature models, but different
  probabilities $P_\mathrm{random}$ in the \textit{runs test} for
  randomness. Best fits are reached for
  $\dot{m}=5-10\,\mathrm{g}\,\mathrm{cm}^{-2}\,\mathrm{s}^{-1}$ in combination
  with the single-temperature black body at $P_\mathrm{random}=68\,\%$, and
  for $\dot{m}=0.1\,\mathrm{g}\,\mathrm{cm}^{-2}\,\mathrm{s}^{-1}$ in
  combination with the multitemperature black bodies at a high
  $P_\mathrm{random}=89\,\%$.  This multi-black body and \textsc{mekal}
  best-fit model results in essentially the same black-body temperatures as
  described in Sect.~\ref{sec:softspec} and a wide range of plasma
  temperatures between 2.6 and 62\,keV with a flux-weighted mean of
  $kT_\mathrm{plasma}=13.3^{+8.3}_{-6.1}\,\mathrm{keV}$ at a total unabsorbed
  flux of $F_\mathrm{bol,column}=1.3^{+8.3}_{-0.6}\times
  10^{-12}\,\mathrm{erg\,cm}^{-2}\,\mathrm{s}^{-1}$ and higher intrinsic
  absorption of $N_\textsc{H,pcfabs}=5.2^{+13.3}_{-2.9}\times
  10^{23}\,\mathrm{cm}^{-2}$.

%
%

  \begin{figure*}
    \includegraphics[width=18.2cm]{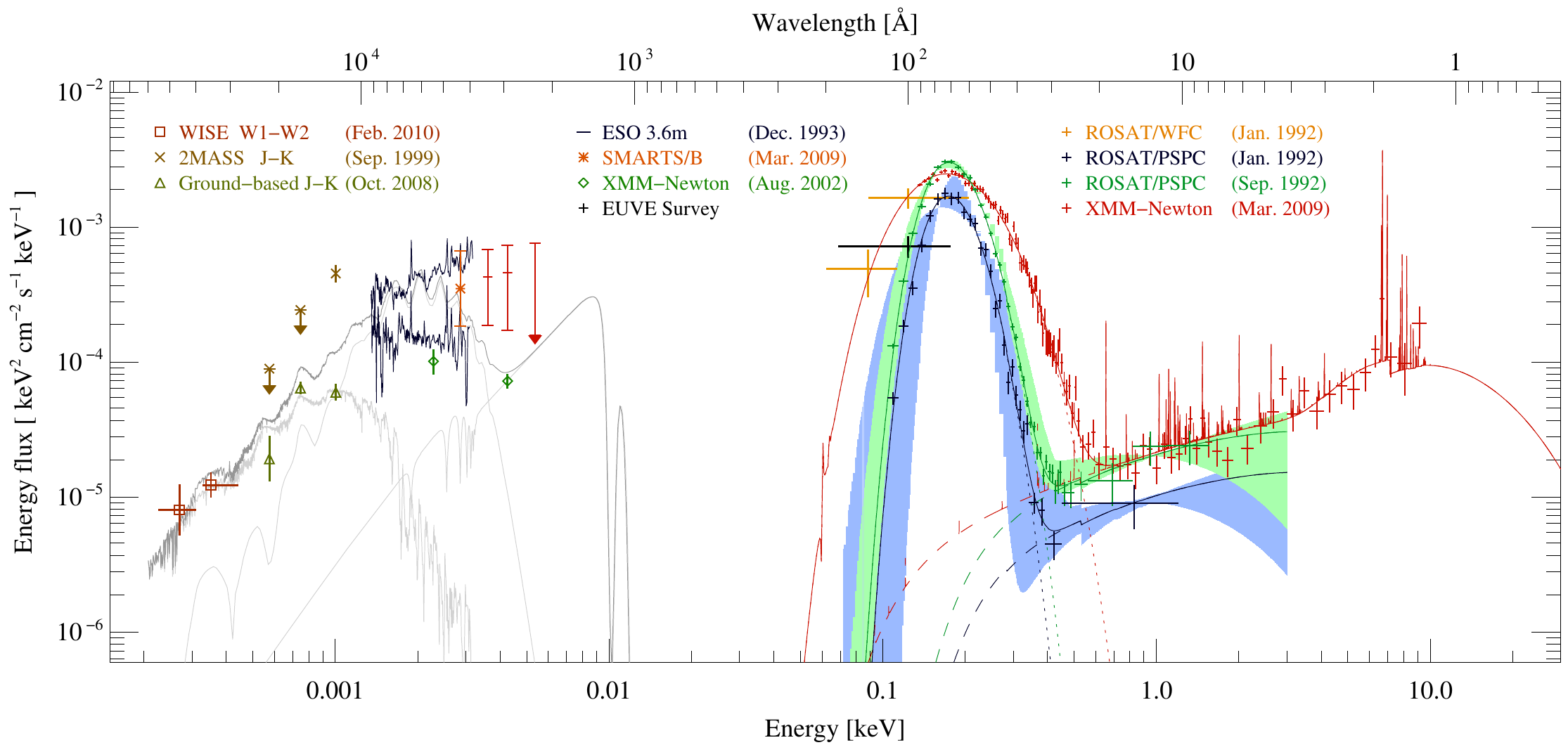}
    \caption{Spectral energy distribution of RS~Cae during high and low states
      of accretion from 1992 to 2010: archival data of various missions from
      the infrared to the X-ray bands and our 2009 observations. The optical
      and UV spectroscopic and photometric data are given as orbital minimum
      and maximum, and 2MASS $H$- and $K$-band fluxes as upper limits. The
      X-ray spectra are shown with the models (solid lines) and their
      components (dashed and dotted). The shaded areas mark the confidence
      ranges of the ROSAT temperatures. The gray lines represent stellar and
      cyclotron model spectra (light gray), and their sum (dark gray, details
      are given in the text).}
    \label{fig:sed}%
  \end{figure*}

%
%
\section{Towards a combined SED model}
\label{sec:disc_sedmodeling}

\subsection{The spectral energy distribution}
\label{sec:disc_sed}


  Aiming for a physically realistic description of the whole system, we
  inspected the spectral energy distribution (SED) of RS~Cae and studied the
  contributions of the individual system components over the whole energy
  range.
%
%
  Figure~\ref{fig:sed} shows the SED in a synopsis of high- and low state
  observations between 1992 and 2010, including our SMARTS and XMM-Newton
  data. The ROSAT/WFC count rates of \citet{pye:95} have been converted into
  fluxes using the conversion factors according to \citet{hodgkin:94}, and the
  EUVE Survey data according to \citet{bowyer:96}. From the first XMM-Newton
  observation in 2002, obtained during a low state of the system, the
  Optical-Monitor measurements are included in the plot, while RS~Cae was too
  faint to be detected by the EPIC and RGS instruments \citep[observation ID
    0109464301, ][]{ramsay:04lowstates}. Since it has low infrared and optical
  fluxes even during high states of accretion, the WISE data are low
  signal-to-noise data taken from the reject catalog, and the 2MASS data are
  ``B'' quality data (detection valid to $80\,\%-90\,\%$). The survey data
  (WISE, 2MASS, EUVE, WFC) are snapshots of RS~Cae at unknown orbital phase,
  while the ESO, SMARTS, and XMM-Newton/OM fluxes are given as orbital minimum
  and maximum during high states and as orbital means during low states. The
  X-ray spectra are shown unfolded with their respective best-fit models: the
  XMM-Newton/EPIC data with the absorbed black-body plus \textsc{mekal} fit as
  described in Sect.~\ref{sec:spectra}, and the archival PSPC data with
  black-body plus bremsstrahlung fits. With the bremsstrahlung temperature
  being fixed to $kT_\textsc{brems}:=5\,\mathrm{keV}$, they yield
  $N_\textsc{H,tbnew}=1.5^{+1.1}_{-0.8}\times 10^{20}\,\mathrm{cm}^{-2}$,
  $kT_\textsc{bbody}=17.9^{+9.6}_{-8.8}\,\mathrm{eV}$ for the 1992 January
  data and $N_\textsc{H,tbnew}=1.3^{+0.6}_{-0.5}\times
  10^{20}\,\mathrm{cm}^{-2}$,
  $kT_\textsc{bbody}=19.6^{+6.3}_{-4.6}\,\mathrm{eV}$ for the 1992 September
  data. These values are mostly independent of the bremsstrahlung temperature,
  varying only on a sub-percent level within an $1-20\,\mathrm{keV}$ interval
  of $kT_\textsc{brems}$, but poorly constrained.


  We compare the observed data points with models for the different system
  components and theoretical expectations.

\subsection{Spectral models and parameters}
\label{sec:disc_models}

  In addition to our X-ray spectral fits, we apply three models for the cooler
  emission in the infrared to the ultraviolet range:
  \begin{description}
    \item[\textit{(i)}] emission from the secondary M-dwarf atmosphere, 
    \item[\textit{(ii)}] cyclotron emission from the cooling post-shock
      accretion flow, and
    \item[\textit{(iii)}] emission from the unheated white-dwarf atmosphere.
  \end{description}
  The spectral contribution of the preshock
  accretion stream is not considered in the plot.  We include
  \begin{description}
    \item[\textit{(i)}] a PHOENIX model \citep{hauschildt:99}, as
      employed by \citet{heller:11} in their analyses of white-dwarf
      M-star binaries. It depends on temperature, surface gravity, and
      chemical composition of the M-dwarf;
    \item[\textit{(ii)}] a cyclotron model spectrum by
      \citet{fischer:01}. It depends on the magnetic
      field strength, the mass of the white dwarf, and the local
      mass-flow densities in the accretion column; and
    \item[\textit{(iii)}] a non-LTE model atmosphere of a hydrogen-helium
      white dwarf, using routines of the T\"ubingen NLTE Model-Atmosphere
      Package \citep{werner:86,werner:99}. It depends on temperature, surface
      gravity (or: mass and age), and chemical composition of the white dwarf.
  \end{description}
  In the following, we describe the parameter set we use for the three
  models.

  Parameters that are constant in time are taken from the literature. For the
  system geometry, we assume a binary inclination $i\sim 60\degr$ and a
  colatitude $\beta\sim 25\degr$ of the accretion region on the white dwarf,
  following the stream-eclipse scenario of \cite{burwitz:96}. For the
  secondary star (\textit{i}), we estimate the parameters via the empirical
  relations by \citet{knigge:06,knigge:07err} for
  $P_\mathrm{orb}=1.7\,\mathrm{hrs}$: the effective temperature to
  $T_{\mathrm{eff},2}=3\,000\,\mathrm{K}$, the surface gravity to $\log
  g_2=5.0\,[\mathrm{cm\,s}^{-2}]$, and the radius to $R_2=0.166\,R_\odot$. For
  the white dwarf (\textit{ii} and \textit{iii}), we use the magnetic field
  strength $B=36\,\mathrm{MG}$ of \citet{burwitz:96} and typical values of
  white dwarfs in polars as reviewed, for example, by
  \citet{sion:99,kawka:07,townsley:09}:
  $T_\mathrm{eff,WD}=15\,000\,\mathrm{K}$, $\log
  g_\mathrm{WD}=8.0\,[\mathrm{cm\,s}^{-2}]$, and $M_\mathrm{WD}=0.6\,M_\odot$,
  corresponding to a radius of about $R_\mathrm{WD}\sim 0.012\,R_\odot$
  \citep{koester:86}. In all models, we assume solar element abundances.

  The distance to the system cannot be derived directly from the optical
  spectra, because they lack the features of the secondary star. The models
  (\textit{i} to \textit{iii}) are fully consistent with the ultraviolet,
  optical, and infrared measurements shown in Fig.~\ref{fig:sed} if we scale
  them to a distance of 750\,pc. At this value, the M-star model coincides
  with the ground-based $JHK$ measurements, which serve as an upper limit for
  the stellar contribution and, thus, as lower limit for the distance
  estimate. The low $J$-to-$H$ magnitude ratio supports the interpretation
  that the 2008 data are M-star-dominated and represent a low state of
  accretion, while the steeper 2MASS measurements include significant
  cyclotron emission during a high state. The 750\,pc agree with the lower
  limit of 440\,pc given by \citet{burwitz:96} and with the determination of
  $880^{+300}_{-220}\,\mathrm{pc}$ of \citet{pretorius:13}. It mainly depends
  on the stellar radii of the white dwarf and the M-star. A distance less than
  750\,pc would require smaller radii for both stars in order not to exceed
  the data; a longer distance would require larger stellar radii to match the
  observed data. These radii would conflict with the empirical values of
  \citet{knigge:06}.

  Time-variable parameters are mass-flow rates, accretion temperatures, and
  emitting areas. They have to be fitted per observational epoch. X-ray
  temperatures, emitting surface areas, and column densities are derived from
  the X-ray spectral fits (Sects.~\ref{sec:spectra} and
  \ref{sec:disc_sed}). The mass-flow density in the accretion column is a free
  parameter of the models of \citet{fischer:01}. We estimate it by scaling the
  models to approximately match the shape of the ESO spectra observed in 1993:
  two local mass-flow densities in the accretion column,
  $\dot{m}_1=0.01\,\mathrm{g\,cm}^{-2}\,\mathrm{s}^{-1}$ with a column base
  area $A_1=10^{16}\,\mathrm{cm}^2$ and
  $\dot{m}_2=0.1\,\mathrm{g\,cm}^{-2}\,\mathrm{s}^{-1}$ with $A_2=1.3\times
  10^{15}\,\mathrm{cm}^2$. The bremsstrahlung flux of the \citet{fischer:01}
  models are also in accordance with the hard X-ray fluxes in the
  XMM-Newton/EPIC observation and their upper limits in the ROSAT/PSPC
  observations, when using a partially covering absorber on the same order as
  in the spectral fits in Sect.~\ref{sec:spectra}.

\subsection{Application to the multiband light curves}
\label{sec:disc_lcsim}

  From the combined spectral models, we derived model light curves in optical,
  ultraviolet, and soft X-ray bands, aiming at a physical interpretation of
  the light curves and a consistency check for our models. This is the first
  effort to model the multiwavelength light curves of a polar, combining
  photometric with spectroscopic information during high and low states of
  accretion. For simulating the low-energy light curves, we consider cyclotron
  emission from the accretion column, accretion-stream emission, and
  white-dwarf emission. The contribution of the secondary star is negligible
  in the wave bands of our photometry (cf.\ Fig.~\ref{fig:sed}).

  We calculate the cyclotron light curves from the phase-resolved column
  spectra (model \textit{ii} in Sect.~\ref{sec:disc_models}), folding them
  with the transmission curves of the Johnson-Cousins and XMM-Newton/OM
  filters. They show a double-humped structure that has been observed and
  attributed to cyclotron beaming in other polars, as in AM~Her
  \citep{gaensicke:01}, \object{AR~UMa} \citep{howell:01}, and \object{HU~Aqr}
  \citep{schwope:03}. In our models, the phasing of the deepest minimum
  changes from $\varphi_\mathrm{X-ray}=0.0$ in the infrared, $U$, and $B$
  light curves to $\varphi_\mathrm{X-ray}=0.5$ in the $VRI$ light curves. The
  white-dwarf fluxes are given by the atmosphere model (\textit{iii}) of
  Sect.\ \ref{sec:disc_models}. The flux modulation of the preshock accretion
  stream with the orbital phase is calculated within a 3D binary model for a
  constant temperature along the stream \citep{staude:01}, which means the
  same \textit{relative} stream-flux modulation in all filters. The
  \textit{absolute} stream flux is guessed per filter from the light curve
  minima as measured flux minus cyclotron and white-dwarf flux.

\begin{figure}
    \centering
    \includegraphics[width=7.2cm]{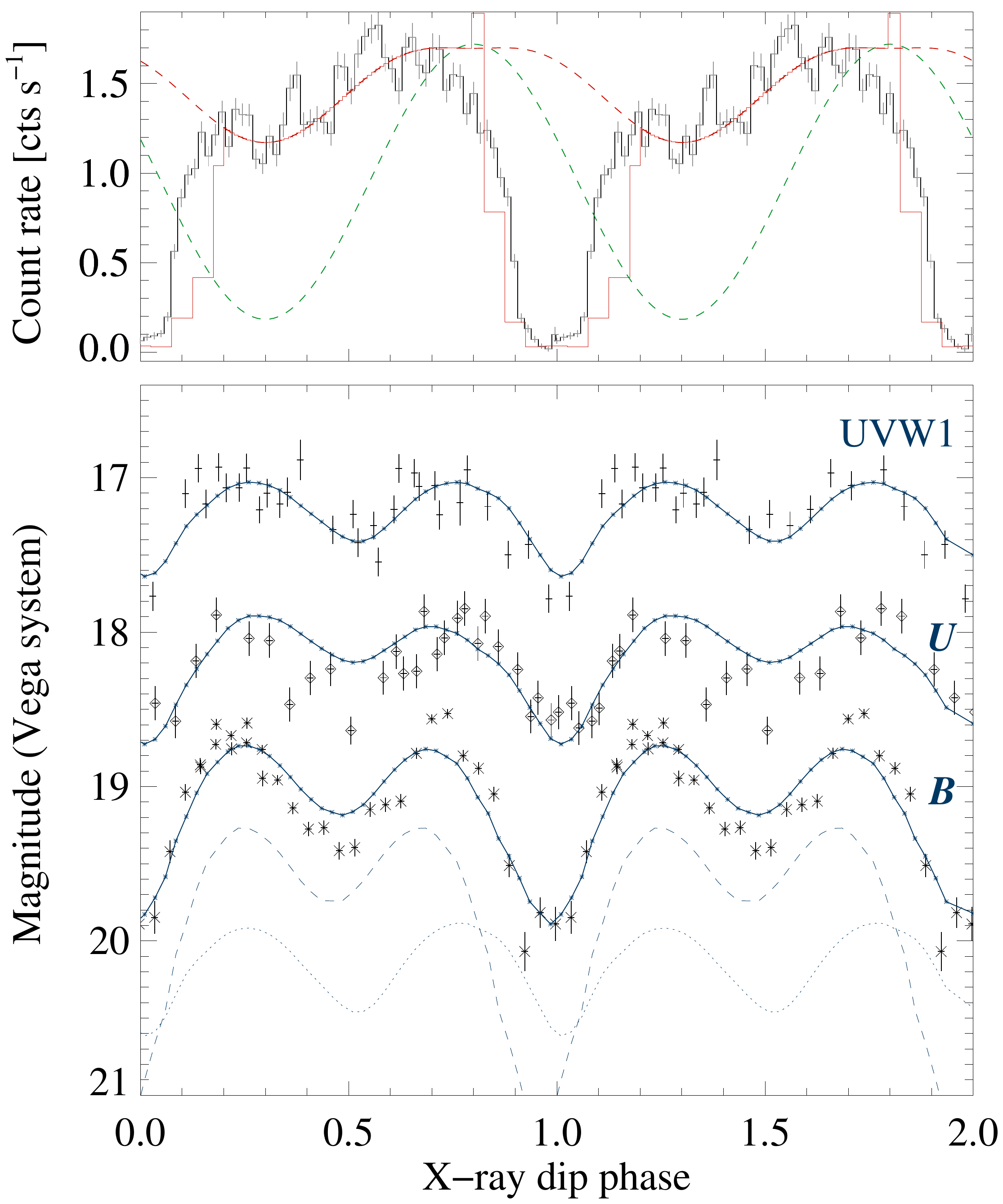}
    \caption{Observed (2009) and simulated light curves of
      RS~Cae. \textbf{Upper panel:} Phase-averaged soft X-ray light curve in
      the black-body-dominated $0.1\,\mathrm{keV} \leq E \leq
      0.5\,\mathrm{keV}$ band plus simulated light curves of a flat circular
      (dashed green) and a cylindrical (red) accretion region. The solid curve
      includes the absorption-dip fit described in the text. EPIC/pn time bins
      are 100\,s. \textbf{Lower panel:} Optical monitor UVW1 filter (shifted
      by $-0.7\,\mathrm{mag}$), $U$ filter, and SMARTS $B$-band. The dashed
      and dotted lines represent the cyclotron and the accretion-stream
      contribution to the $B$-band simulation, respectively. OM time bins are
      300\,s.}
    \label{fig:simlcs}%
  \end{figure}

  To combine the different light curve components, we need their relative
  phasing, i.\,e.,\ information on the system geometry. The spectroscopic
  ephemeris is not accurate enough to be extrapolated to 2009 and to
  independently determine the orbital phasing. We therefore estimate the phase
  shifts between the components directly from the observed light curves: the
  shift between X-ray (dip) phase and cyclotron (magnetic) phase from the
  primary optical light-curve minimum to
  $\varphi_\mathrm{X-ray}-\varphi_\mathrm{cycl}\sim -0.04$, and the shift
  between cyclotron and stream flux via the color dependence of the secondary
  minimum and the asymmetric shape of the optical and UV light curves to
  $\varphi_\mathrm{cycl}-\varphi_\mathrm{stream}\sim
  -0.055$. Figure~\ref{fig:simlcs} shows our 2009 UVW1, $U$, and $B$
  observations, along with the final simulations.

  In addition, we model the soft X-ray light curves as orbital projections of
  emitting black-body surface areas, to which the measured black-body flux is
  proportional. We employ the same binary inclination $i\sim 60\degr$ and
  colatitude $\beta\sim 25\degr$ of the accretion region as in
  Sect.~\ref{sec:disc_models}. We start with the simplest approach of a flat
  circular accretion region (Fig.~\ref{fig:simlcs}), which results in a
  significantly higher light-curve amplitude than measured for RS~Cae. The
  observed amplitude could only be explained by a flat emission region if both
  inclination $i$ and colatitude $\beta$ were on the order of $25\degr$, which
  is inconsistent with the optical data of \citet{burwitz:96}. In fact, the
  soft X-ray and EUV emitting regions are expected to be bulging and
  arc-shaped rather than flat and circular, corresponding to extended,
  arc-shaped bases of the post-shock accretion columns
  \citep[cf.][]{cropper:89,ferrario:90,potter:97}.  Observationally, evidence
  of more complexity was found, for example, by
  \citet{vennes:95,sirk:98,gaensicke:98,szkody:99}.  We tested for a
  three-dimensional shape of the accretion region of RS~Cae with the
  projection of a cylindrical emission region. The flux from an extended
  cylinder with a height of 1.25 times its diameter reproduces the soft
  EPIC/pn light curve reasonably well (Fig.~\ref{fig:simlcs}). We derive a
  phase shift of about $\varphi_\mathrm{X-ray} - \varphi_\mathrm{softX}\sim
  -0.2$ from the light curves, where $\varphi_\mathrm{softX}$ denotes maximum
  soft X-ray flux, i.e.,\ maximum visibility of the accretion region. This
  shift means that the centers of accretion region and accretion column might
  be offset.

  To account for the stream-absorption dip in our simulations of the soft
  X-ray light curves, we use phase-resolved spectral models of the dip phase.
  We extract EPIC/pn spectra at energies below 0.5\,keV in a $\pm 0.225$ phase
  range around the dip center with a phase resolution of $\Delta\varphi=0.05$
  and fit them simultaneously with absorbed black-body models. Since the
  $\Delta\varphi=0.05$ spectra during dip phase comprise only a few bins, we
  repeat the fit with spectra extracted for overlapping $\Delta\varphi=0.1$
  intervals, centered at $\varphi_\mathrm{X-ray}=0.00, 0.05, 0.10,$ etc.,
  similar to the approach that \citet{girish:07} use to fit the iron lines of
  AM~Her. We use identical black-body temperatures for all phase intervals and
  couple the normalizations (i.e., emitting surface areas) according to the
  projected areas of the cylindrical emission region. The hydrogen absorption
  on the line of sight increases in our fits up to $N_\textsc{H,tbabs}\sim
  8\times 10^{22}\,\mathrm{cm}^{-2}$, including interstellar absorption,
  during the dip phase. With a reduced $\chi^2_\mathrm{red}=1.5$ of the
  simultaneous black-body fit, modeling the emission region as a
  three-dimensional cylinder is significantly more appropriate than as a flat
  circle, but still lacking. We convert the $\Delta\varphi=0.05$ model fluxes
  in the $0.1-0.5\,\mathrm{keV}$ interval to absorption factors by dividing
  the absorbed fluxes by the fluxes for a constant interstellar (nondip)
  absorption of $N_\mathrm{H}=2.1\times 10^{19}\,\mathrm{cm}^{-2}$ and
  multiply the simulated X-ray light curves by them. The results are shown in
  the upper panel of Fig.~\ref{fig:simlcs} and give a reasonable description
  of the absorption dip. Deficits of the model manifest themselves in
  particular between $\varphi_\mathrm{X-ray}=0.1$ and 0.2, where the count
  rate is underestimated because of the simplifications made in modeling the
  emitting areas and the phase shifts.


\subsection{Results}
\label{sec:disc_sedres}

  With our SED and light curve models, we separate the contributions of the
  different system components to the spectral energy distribution and derive a
  new lower limit estimate of the distance to RS~Cae. In the infrared range, a
  large part of the total flux can be attributed to the M-type secondary, plus
  cyclotron emission during high states of accretion. The cyclotron component
  is dominating the stellar emission in the optical and near-infrared
  range. In particular, large parts of the $B$-band modulation can be
  attributed to cyclotron emission (Fig.~\ref{fig:simlcs}). The color
  dependence of the light-curve minima is explained by the decreasing
  contribution of the cyclotron flux and increasing contribution of
  accretion-stream flux from the optical toward the ultraviolet. The cyclotron
  model spectra reproduce the high-state data in 1993 and 2009, and the 2MASS
  data indicate that even higher infrared fluxes may be reached at other
  epochs. The unheated white dwarf contributes little to the high-state flux,
  with the low-state UV flux serving as upper limit for the white dwarf.

\section{The energy balance of RS~Cae}
\label{sec:disc_energy}

  Since RS~Cae was discovered at a hardness ratio close to $-1.0$ during the
  ROSAT All-Sky Survey \citep{thomas:98}, it has been considered to be one of
  the soft X-ray dominated polars. In these systems, the soft X-ray luminosity
  exceeds the $50\,\%$ of the total luminosity predicted by the standard model
  of accretion \citep{king:79,lamb:79}. Former ROSAT results are biased
  towards soft energies, since the energy range was limited to
  $0.1-2.5\,\mathrm{keV}$, and have to be verified over a broader energy
  range. With our fits to the XMM-Newton spectra, we investigate the energy
  balance of RS~Cae on the basis of data in the $0.1-0.5\,\mathrm{keV}$ and
  $0.5-10.0\,\mathrm{keV}$ bands, and with our SED models on the basis of the
  optical and ultraviolet measurements.

  \subsection{XMM-Newton fluxes and X-ray flux ratios}

  From the absorbed single-temperature model described in
  Sect.~\ref{sec:spectra}, we have derived bolometric fluxes of
  $F_\mathrm{bol}(\textsc{bbody})=7.9^{+1.0}_{-0.8} \times
  10^{-12}\,\mathrm{erg\,cm}^{-2}\,\mathrm{s}^{-1}$ and
  $F_\mathrm{bol}(\textsc{mekal})=7.3^{+0.5}_{-0.5} \times
  10^{-13}\,\mathrm{erg\,cm}^{-2}\,\mathrm{s}^{-1}$, and a soft-to-hard flux
  ratio of $F_{\mathrm{bol,}\textsc{bbody}} /
  F_{\mathrm{bol,}\textsc{mekal}}=10.8^{+1.4}_{-1.0}$ during non-dip
  phases. For the cyclotron component, we estimate a bolometric flux of about
  $F_\mathrm{bol,cycl}\sim 5\times
  10^{-13}\,\mathrm{erg\,cm}^{-2}\,\mathrm{s}^{-1}$ from the accretion-column
  model presented in Sect.~\ref{sec:disc_models}.

  The flux values strongly depend on the choice of the spectral model
  (cf.\ Table~\ref{tab:xspecfits}). \citet{ramsay:04ebalance} present flux and
  luminosity ratios based on single-temperature black-body and
  multitemperature column models. Multitemperature models typically result in
  higher bolometric fluxes: Since the flux in the instrumental energy window
  is scaled to the observed flux, the fluxes toward the (unobserved) lower and
  higher energies are raised compared to a single-temperature continuum owing
  to the broader temperature range (cf.\ small panel in
  Fig.~\ref{fig:spectra}). Our fits indicate that the spectra reflect the
  multitemperature nature both of accretion region and accretion column, while
  higher spectral resolution and sensitivity would be needed to derive the
  parameter distributions directly from the observed data. In the combined
  multicomponent fit with predefined temperature distributions, the bolometric
  fluxes of both the multi-\textsc{bbodyrad} and the multi-\textsc{mekal}
  component increase by a factor of 1.5, respectively, leaving the flux ratio
  essentially unchanged.

  The soft X-ray excess of polars that was measured from ROSAT data increases
  with magnetic field strength \citep{beuermann:94,ramsay:94}. The excess in
  RS~Cae is comparable to the XMM-Newton results for polars of similar field
  strengths: \object{EK~UMa} ($B\sim 35\,\mathrm{MG}$) with a moderate
  luminosity ratio of six for single-temperature models during a short 5\,ks
  exposure \citep{ramsay:04ebalance}; AI~Tri ($B\sim 38\,\mathrm{MG}$) with
  moderate to high flux ratios of 6 to at least 70, depending on the accretion
  state \citep{traulsen:10}; HU~Aqr ($B\sim 35\,\mathrm{MG}$) with low flux
  ratios during a low state of accretion, balanced fluxes or a slight excess
  during an intermediate state, and a strong excess during a high state
  \citep{schwarz:09}.

  \begin{table}
    \caption{Emitting black-body areas and bolometric fluxes of RS~Cae during
      the high-state observations by ROSAT in 1992 and XMM-Newton in 2009,
      together with their 90\,\% confidence intervals.}
    \label{tab:fluxes}\centering
    \begin{tabular}{lcc}
      \hline\hline\\[-2ex]
       & ROSAT 1992 & XMM-Newton 2009 \\ \hline\\[-1.5ex]
     $A_\textsc{bbody}$ 
         &  $\sim 10^{16}$           
         & \multirow{2}{*}{$5.1^{+1.1}_{-0.8}\times 10^{13}$} \\
     $~~~[\mathrm{cm}^{-2}]$
         &  $([0.3, 9]\times 10^{16})$  
         &  \\[.8ex]
     $F_\textsc{bbody}$ 
         &  $\sim 2\times{10^{-10}}$ 
         & \multirow{2}{*}{$7.9^{+1.0}_{-0.8}\times 10^{-12}$} \\
     $~~~[\mathrm{erg\,cm}^{-2}\,\mathrm{s}^{-1}]$ 
         &  $(\gtrsim 5\times 10^{-11})$ 
         & \\[.8ex]
     $F_\textsc{brems}~~(5\,\mathrm{keV})$
         &  $9\times 10^{-14}$        
         & \multirow{2}{*}{$1.6^{+0.1}_{-0.1}\times 10^{-13}$} \\
     $~~~[\mathrm{erg\,cm}^{-2}\,\mathrm{s}^{-1}]$
         &  $([6, 12]\times 10^{-14})$  
         & \\
      \hline
    \end{tabular}
  \end{table}

  \subsection{XMM-Newton and optical luminosities}

  We convert the bolometric fluxes to luminosities as $L=\eta\pi Fd^2$, where
  $\eta$ denotes the geometric correction factor. Typically, for the hard
  component a factor of $3\pi$ (considering reflection effects) up to $4\pi$
  is chosen, while \citet{king:87} emphasize that a factor of $2\pi$ for
  column emission into one half space would be more appropriate.  For the soft
  emission of a flat accretion region, a factor of $2\pi$ or
  $\pi\mathrm{cos}^{-1}\vartheta$, depending on the viewing angle $\vartheta$,
  is used, and of $4\pi$ for blobby accretion when accretion mounds form at
  the impact area due to the heating of the photosphere
  \citep{hameury:88,beuermann:89}. Having shown the evidence of a
  three-dimensional structure of the accretion region in RS~Cae, we choose the
  same geometric factor of $4\pi$ for the X-ray soft and for the X-ray hard
  component, and $2\pi$ for the cyclotron component. For the multitemperature
  model, we thus obtain $L_\mathrm{softX}=7.7\times
  10^{32}\,\mathrm{erg\,s}^{-1}$, $L_\mathrm{hardX}=7.3\times
  10^{31}\,\mathrm{erg\,s}^{-1}$, $L_\mathrm{cycl}\sim1.7\times
  10^{31}\,\mathrm{erg\,s}^{-1}$ for a distance of $d=750\,\mathrm{pc}$, and
  soft-to-hard luminosity ratios on the same order of 10 as the flux ratios.

  The accretion-induced luminosities of soft X-ray, hard X-ray, and cyclotron
  component add to $L_\mathrm{accr}=8.6\times 10^{32}\,\mathrm{erg\,s}^{-1}$,
  without the (unknown) contribution of the preshock accretion stream. For
  a white-dwarf mass of $M_\mathrm{WD}=0.6\,\mathrm{M}_\odot$ and radius of
  $R_\mathrm{WD}\sim 0.012\,R_\odot$, this corresponds to a mass-accretion rate
  on the order of
  $\dot{M}=L_\mathrm{accr}\,R_\mathrm{WD}/(G\,M_\mathrm{WD})\sim 
  10^{-10}\,\mathrm{M}_\odot\,\mathrm{yr}^{-1}$, within the typical range of 
  high-state accretion rates of polars.

  \subsection{Long-term characteristics of soft and hard emission}

  The SED plot in Fig.~\ref{fig:sed} shows the distinct soft X-ray components
  in the ROSAT and XMM-Newton observations. They changed significantly from
  1992 to 2009, seen in the emitting black-body areas, which represent the
  area at the mean temperature of the accretion region, and, correspondingly,
  in the (bolometric) fluxes of the black-body fits (Table~\ref{tab:fluxes}):
  The soft X-ray fluxes during the ROSAT observation of September 1992 are by
  a factor of at least 5.5 higher than during the 2009 XMM-Newton
  observation. The hard X-ray and cyclotron fluxes, on the other hand, are on
  the same order of magnitude during the 2009 observations as in 1992/93. Both
  hard X-ray and cyclotron emission are assumed to arise from the accretion
  column, so the similar flux levels may indicate similar physical properties
  of the column at both epochs. Although the ROSAT fit results are poorly
  constrained, they show clearly that the long-term characteristics of soft
  and hard X-ray component are not correlated. \citet{gaensicke:95}
  demonstrated for AM~Her that the bremsstrahlung and cyclotron flux are
  balanced by the (reprocessed) UV flux both during high and low states of
  accretion, independently of the soft X-ray component, which arises from
  blobby accretion. Correspondingly, the soft stages of RS~Cae and its
  decoupled soft and hard X-ray fluxes may indicate inhomogeneous accretion
  processes.

%
%
\section{Summary and conclusions}

  Our pointed XMM-Newton observation of RS~Cae, covering 4.5 orbital cycles,
  provides the first opportunity to investigate its energy balance on the
  basis of data at energies up to 10\,keV. RS~Cae was clearly in a high state
  of accretion at the epoch of our observations. Archival infrared and X-ray
  data indicated that still higher and potentially softer states might be
  possible. Almost $97\,\%$ of the EPIC/pn photons were detected in the soft
  range, yielding hardness ratios close to $-1$ in the X-ray light curves. The
  light curves gave evidence of a three-dimensional shape of the accretion
  region and inhomogeneous accretion events, as is typical of soft polars. We
  identified the sharp, recurring light-curve dips as stream absorption and
  used them to derive a photometric period of 0.0709(3)\,days, confirming the
  preferred period of \citet{burwitz:96} and placing RS~Cae among the
  short-period polars. The energy dependence of the width of the dip, becoming
  broader towards higher energies, indicated different spatial extents of the
  X-ray emitting regions.

  Using SED modeling, we consistently connected the multiwavelength spectra
  and light curves of the different system components. SEDs constructed of
  nonsimultaneous observations provide insight into long-term behavior and
  accretion-stage changes in the system, but have a limited ability to model
  the SED. This successful approach shows the potential of SED modeling and
  simultaneous multiwavelength observations of polars.

  Using single- and multitemperature fits to the EPIC spectra, we find a soft
  X-ray excess with soft-to-hard luminosity ratios of about ten. Our
  multitemperature spectral models give physically plausible descriptions of
  the structure of the accretion region and column and the respective X-ray
  and low-energy spectra and light curves. They result in the same
  soft-to-hard flux ratios as the single-temperature models.

  Up to now, we have studied three systems with a clear soft X-ray excess
  during (intermediate) high states of accretion in the ROSAT and in our
  pointed XMM-Newton observations: AI~Tri \citep{traulsen:10}, QS~Tel
  \citep{traulsen:11}, and RS~Cae. Their soft X-ray luminosity correlates with
  their accretion state and can change drastically on time scales as short as
  days, in agreement with the dependence of the soft-to-hard luminosity ratio
  on the accretion state as described by \citet{ramsay:04ebalance}.  The three
  systems augment the number of polars whose ROSAT-detected soft X-ray excess
  could be confirmed over the broader energy range of XMM-Newton. Which
  fraction of AM~Her-type systems actually shows a soft X-ray excess is hard
  to determine from the currently available data.  One limiting factor has
  been instrumental biases. With higher sensitivity in particular in the hard
  X-ray regime, an increasing number of X-ray hard and low accretion rate
  polars are being detected. Another limiting factor is observational
  biases. In untriggered observations, a substantial number of polars are
  caught during low states, so their soft stages would be missed. Considering
  the selection effects and our observational results, we expect that the
  fraction of 25\,\% X-ray soft polars among ROSAT detections might
  overestimate the actual number, but that the soft systems form a significant
  group among the polars.

%
%
\begin{acknowledgements}

  This research has been supported by the DLR under project numbers
  50\,OR\,0501, 50\,OR\,0807, and 50\,OR\,1011. We thank Ren\'e Heller for
  providing the M-star spectrum shown in Fig.~\ref{fig:sed} and Karleyne Silva
  for calculating accretion-column light curves and for fruitful
  discussions. FWM's access to the SMARTS observatory is supported in part by
  a NASA grant NNX10AE51G to Stony Brook University.

  Figure~\ref{fig:sed} includes data from the High Energy Astrophysics Science
  Archive Research Center (HEASARC), provided by NASA's Goddard Space Flight
  Center; data products from the Two Micron All Sky Survey, which is a joint
  project of the University of Massachusetts and the Infrared Processing and
  Analysis Center/California Institute of Technology, funded by the National
  Aeronautics and Space Administration and the National Science Foundation;
  and data products from the Wide-field Infrared Survey Explorer, which is a
  joint project of the University of California, Los Angeles, and the Jet
  Propulsion Laboratory/California Institute of Technology, funded by the
  National Aeronautics and Space Administration.
\end{acknowledgements}

%
%
\bibliographystyle{aa}
\bibliography{21383}

\begin{thebibliography}{74}
\expandafter\ifx\csname natexlab\endcsname\relax\def\natexlab#1{#1}\fi

\bibitem[{{Arnaud}(1996)}]{arnaud:96}
{Arnaud}, K.~A. 1996, in ASP Conf.\ Ser., Vol. 101, Astronomical Data Analysis
  Software and Systems V, ed. G.~H. {Jacoby} \& J.~{Barnes}, 17

\bibitem[{{Asplund} {et~al.}(2009){Asplund}, {Grevesse}, {Sauval}, \&
  {Scott}}]{asplund:09}
{Asplund}, M., {Grevesse}, N., {Sauval}, A.~J., \& {Scott}, P. 2009, \araa, 47,
  481

\bibitem[{{Beardmore} {et~al.}(1995){Beardmore}, {Done}, {Osborne}, \&
  {Ishida}}]{beardmore:95}
{Beardmore}, A.~P., {Done}, C., {Osborne}, J.~P., \& {Ishida}, M. 1995, \mnras,
  272, 749

\bibitem[{{Bernardini} {et~al.}(2012){Bernardini}, {de Martino}, {Falanga},
  {Mukai}, {Matt}, {Bonnet-Bidaud}, {Masetti}, \& {Mouchet}}]{bernardini:12}
{Bernardini}, F., {de Martino}, D., {Falanga}, M., {et~al.} 2012, \aap, 542,
  A22

\bibitem[{{Beuermann} \& {Burwitz}(1995)}]{beuermann:95}
{Beuermann}, K. \& {Burwitz}, V. 1995, in ASP Conf.\ Ser., Vol.~85, Magnetic
  Cataclysmic Variables, ed. D.~A.~H. {Buckley} \& B.~{Warner}, 99

\bibitem[{{Beuermann} {et~al.}(2012){Beuermann}, {Burwitz}, \&
  {Reinsch}}]{beuermann:12}
{Beuermann}, K., {Burwitz}, V., \& {Reinsch}, K. 2012, \aap, 543, A41

\bibitem[{{Beuermann} \& {Schwope}(1989)}]{beuermann:89}
{Beuermann}, K. \& {Schwope}, A.~D. 1989, \aap, 223, 179

\bibitem[{{Beuermann} \& {Schwope}(1994)}]{beuermann:94}
{Beuermann}, K. \& {Schwope}, A.~D. 1994, in ASP Conf.\ Ser., Vol.~56,
  Interacting Binary Stars, ed. A.~W. {Shafter}, 119

\bibitem[{{Beuermann} {et~al.}(1999){Beuermann}, {Thomas}, {Reinsch},
  {Schwope}, {Tr{\"u}mper}, \& {Voges}}]{beuermann:99}
{Beuermann}, K., {Thomas}, H.-C., {Reinsch}, K., {et~al.} 1999, \aap, 347, 47

\bibitem[{{Bowyer} {et~al.}(1996){Bowyer}, {Lampton}, {Lewis}, {Wu},
  {Jelinsky}, \& {Malina}}]{bowyer:96}
{Bowyer}, S., {Lampton}, M., {Lewis}, J., {et~al.} 1996, \apjs, 102, 129

\bibitem[{{Bowyer} {et~al.}(1994){Bowyer}, {Lieu}, {Lampton}, {Lewis}, {Wu},
  {Drake}, \& {Malina}}]{bowyer:94}
{Bowyer}, S., {Lieu}, R., {Lampton}, M., {et~al.} 1994, \apjs, 93, 569

\bibitem[{{Burwitz} {et~al.}(1996){Burwitz}, {Reinsch}, {Schwope}, {Beuermann},
  {Thomas}, \& {Greiner}}]{burwitz:96}
{Burwitz}, V., {Reinsch}, K., {Schwope}, A.~D., {et~al.} 1996, \aap, 305, 507

\bibitem[{{Cropper}(1989)}]{cropper:89}
{Cropper}, M. 1989, \mnras, 236, 935

\bibitem[{{Cropper} {et~al.}(1998){Cropper}, {Ramsay}, \& {Wu}}]{cropper:98}
{Cropper}, M., {Ramsay}, G., \& {Wu}, K. 1998, \mnras, 293, 222

\bibitem[{{Cropper} {et~al.}(1999){Cropper}, {Wu}, {Ramsay}, \&
  {Kocabiyik}}]{cropper:99}
{Cropper}, M., {Wu}, K., {Ramsay}, G., \& {Kocabiyik}, A. 1999, \mnras, 306,
  684

\bibitem[{{Dickey} \& {Lockman}(1990)}]{dickey:90}
{Dickey}, J.~M. \& {Lockman}, F.~J. 1990, \araa, 28, 215

\bibitem[{{Done} \& {Magdziarz}(1998)}]{done:98}
{Done}, C. \& {Magdziarz}, P. 1998, \mnras, 298, 737

\bibitem[{{Dorman} {et~al.}(2003){Dorman}, {Arnaud}, \& {Gordon}}]{dorman:03}
{Dorman}, B., {Arnaud}, K.~A., \& {Gordon}, C.~A. 2003, in Bull.\ Am.\ Astron.\
  Soc., Vol.~35, 641

\bibitem[{{Ferrario} \& {Wickramasinghe}(1990)}]{ferrario:90}
{Ferrario}, L. \& {Wickramasinghe}, D.~T. 1990, \apj, 357, 582

\bibitem[{{Fischer} \& {Beuermann}(2001)}]{fischer:01}
{Fischer}, A. \& {Beuermann}, K. 2001, \aap, 373, 211

\bibitem[{{G{\"a}nsicke} {et~al.}(1995){G{\"a}nsicke}, {Beuermann}, \& {de
  Martino}}]{gaensicke:95}
{G{\"a}nsicke}, B.~T., {Beuermann}, K., \& {de Martino}, D. 1995, \aap, 303,
  127

\bibitem[{{G{\"a}nsicke} {et~al.}(2001){G{\"a}nsicke}, {Fischer}, {Silvotti},
  \& {de Martino}}]{gaensicke:01}
{G{\"a}nsicke}, B.~T., {Fischer}, A., {Silvotti}, R., \& {de Martino}, D. 2001,
  \aap, 372, 557

\bibitem[{{G{\"a}nsicke} {et~al.}(1998){G{\"a}nsicke}, {Hoard}, {Beuermann},
  {Sion}, \& {Szkody}}]{gaensicke:98}
{G{\"a}nsicke}, B.~T., {Hoard}, D.~W., {Beuermann}, K., {Sion}, E.~M., \&
  {Szkody}, P. 1998, \aap, 338, 933

\bibitem[{{Gerke} {et~al.}(2006){Gerke}, {Howell}, \& {Walter}}]{gerke:06}
{Gerke}, J.~R., {Howell}, S.~B., \& {Walter}, F.~M. 2006, \pasp, 118, 678

\bibitem[{{Girish} {et~al.}(2007){Girish}, {Rana}, \& {Singh}}]{girish:07}
{Girish}, V., {Rana}, V.~R., \& {Singh}, K.~P. 2007, \apj, 658, 525

\bibitem[{{Hameury} \& {King}(1988)}]{hameury:88}
{Hameury}, J.~M. \& {King}, A.~R. 1988, \mnras, 235, 433

\bibitem[{{Hauschildt} \& {Baron}(1999)}]{hauschildt:99}
{Hauschildt}, P.~H. \& {Baron}, E. 1999, J.\ Comp.\ Appl.\ Math., 109, 41

\bibitem[{{Heller} {et~al.}(2011){Heller}, {Schwope}, \&
  {{\O}stensen}}]{heller:11}
{Heller}, R., {Schwope}, A.~D., \& {{\O}stensen}, R.~H. 2011, in ASP Conf.\
  Ser., Vol. 447, Evolution of Compact Binaries, ed. L.~{Schmidtobreick}, M.~R.
  {Schreiber}, \& C.~{Tappert}, 177

\bibitem[{{Hodgkin} \& {Pye}(1994)}]{hodgkin:94}
{Hodgkin}, S.~T. \& {Pye}, J.~P. 1994, \mnras, 267, 840

\bibitem[{{Howell} {et~al.}(2001){Howell}, {Gelino}, \& {Harrison}}]{howell:01}
{Howell}, S.~B., {Gelino}, D.~M., \& {Harrison}, T.~E. 2001, \aj, 121, 482

\bibitem[{{Kalberla} {et~al.}(2005){Kalberla}, {Burton}, {Hartmann}, {Arnal},
  {Bajaja}, {Morras}, \& {P{\"o}ppel}}]{kalberla:05}
{Kalberla}, P.~M.~W., {Burton}, W.~B., {Hartmann}, D., {et~al.} 2005, \aap,
  440, 775

\bibitem[{{Kawka} {et~al.}(2007){Kawka}, {Vennes}, {Schmidt}, {Wickramasinghe},
  \& {Koch}}]{kawka:07}
{Kawka}, A., {Vennes}, S., {Schmidt}, G.~D., {Wickramasinghe}, D.~T., \&
  {Koch}, R. 2007, \apj, 654, 499

\bibitem[{{King} \& {Lasota}(1979)}]{king:79}
{King}, A.~R. \& {Lasota}, J.~P. 1979, \mnras, 188, 653

\bibitem[{{King} \& {Watson}(1987)}]{king:87}
{King}, A.~R. \& {Watson}, M.~G. 1987, \mnras, 227, 205

\bibitem[{{Knigge}(2006)}]{knigge:06}
{Knigge}, C. 2006, \mnras, 373, 484

\bibitem[{{Knigge}(2007)}]{knigge:07err}
{Knigge}, C. 2007, \mnras, 382, 1982

\bibitem[{{Koester} \& {Schoenberner}(1986)}]{koester:86}
{Koester}, D. \& {Schoenberner}, D. 1986, \aap, 154, 125

\bibitem[{{Kuijpers} \& {Pringle}(1982)}]{kuijpers:82}
{Kuijpers}, J. \& {Pringle}, J.~E. 1982, \aap, 114, L4

\bibitem[{{Lamb} \& {Masters}(1979)}]{lamb:79}
{Lamb}, D.~Q. \& {Masters}, A.~R. 1979, \apjl, 234, L117

\bibitem[{{Liedahl} {et~al.}(1995){Liedahl}, {Osterheld}, \&
  {Goldstein}}]{liedahl:95}
{Liedahl}, D.~A., {Osterheld}, A.~L., \& {Goldstein}, W.~H. 1995, \apjl, 438,
  L115

\bibitem[{{Magdziarz} \& {Zdziarski}(1995)}]{magdziarz:95}
{Magdziarz}, P. \& {Zdziarski}, A.~A. 1995, \mnras, 273, 837

\bibitem[{{Matt}(2004)}]{matt:04}
{Matt}, G. 2004, \aap, 423, 495

\bibitem[{{McNamara} {et~al.}(2008){McNamara}, {Kuncic}, {Wu}, {Galloway}, \&
  {Cullen}}]{mcnamara:08}
{McNamara}, A.~L., {Kuncic}, Z., {Wu}, K., {Galloway}, D.~K., \& {Cullen},
  J.~G. 2008, \mnras, 383, 962

\bibitem[{{Mewe} {et~al.}(1985){Mewe}, {Gronenschild}, \& van~den
  {Oord}}]{mewe:85}
{Mewe}, R., {Gronenschild}, E.~H.~B.~M., \& van~den {Oord}, G.~H.~J. 1985,
  \aaps, 62, 197

\bibitem[{{Milgrom} \& {Salpeter}(1975)}]{milgrom:75}
{Milgrom}, M. \& {Salpeter}, E.~E. 1975, \apj, 196, 583

\bibitem[{{Nandra} {et~al.}(2007){Nandra}, {O'Neill}, {George}, \&
  {Reeves}}]{nandra:07}
{Nandra}, K., {O'Neill}, P.~M., {George}, I.~M., \& {Reeves}, J.~N. 2007,
  \mnras, 382, 194

\bibitem[{{Potter} {et~al.}(1997){Potter}, {Cropper}, {Mason}, {Hough}, \&
  {Bailey}}]{potter:97}
{Potter}, S.~B., {Cropper}, M., {Mason}, K.~O., {Hough}, J.~H., \& {Bailey},
  J.~A. 1997, \mnras, 285, 82

\bibitem[{{Pounds} {et~al.}(1993){Pounds}, {Allan}, {Barber}, {Barstow},
  {Bertram}, {Branduardi-Raymont}, {Brebner}, {Buckley}, {Bromage}, {Cole},
  {Courtier}, {Cruise}, {Culhane}, {Denby}, {Donoghue}, {Dunford},
  {Georgantopoulos}, {Goodall}, {Gondhalekar}, {Gourlay}, {Harris}, {Hassall},
  {Hellier}, {Hodgkin}, {Jeffries}, {Kellett}, {Kent}, {Lieu}, {Lloyd},
  {McGale}, {Mason}, {Matthews}, {Mittaz}, {Page}, {Pankiewicz}, {Pike},
  {Ponman}, {Puchnarewicz}, {Pye}, {Quenby}, {Ricketts}, {Rosen}, {Sansom},
  {Sembay}, {Sidher}, {Sims}, {Stewart}, {Sumner}, {Vallance}, {Watson},
  {Warwick}, {Wells}, {Willingale}, {Willmore}, {Willoughby}, \&
  {Wonnacott}}]{pounds:93}
{Pounds}, K.~A., {Allan}, D.~J., {Barber}, C., {et~al.} 1993, \mnras, 260, 77

\bibitem[{{Pretorius} {et~al.}(2013){Pretorius}, {Knigge}, \&
  {Schwope}}]{pretorius:13}
{Pretorius}, M.~L., {Knigge}, C., \& {Schwope}, A.~D. 2013, \mnras, 432, 570

\bibitem[{{Pye} {et~al.}(1995){Pye}, {McGale}, {Allan}, {Barber}, {Bertram},
  {Denby}, {Page}, {Ricketts}, {Stewart}, \& {West}}]{pye:95}
{Pye}, J.~P., {McGale}, P.~A., {Allan}, D.~J., {et~al.} 1995, \mnras, 274, 1165

\bibitem[{{Ramsay} \& {Cropper}(2004)}]{ramsay:04ebalance}
{Ramsay}, G. \& {Cropper}, M. 2004, \mnras, 347, 497

\bibitem[{{Ramsay} {et~al.}(2004){Ramsay}, {Cropper}, {Wu}, {Mason},
  {C{\'o}rdova}, \& {Priedhorsky}}]{ramsay:04lowstates}
{Ramsay}, G., {Cropper}, M., {Wu}, K., {et~al.} 2004, \mnras, 350, 1373

\bibitem[{{Ramsay} {et~al.}(1994){Ramsay}, {Mason}, {Cropper}, {Watson}, \&
  {Clayton}}]{ramsay:94}
{Ramsay}, G., {Mason}, K.~O., {Cropper}, M., {Watson}, M.~G., \& {Clayton},
  K.~L. 1994, \mnras, 270, 692

\bibitem[{{Schwarz} {et~al.}(2009){Schwarz}, {Schwope}, {Vogel}, {Dhillon},
  {Marsh}, {Copperwheat}, {Littlefair}, \& {Kanbach}}]{schwarz:09}
{Schwarz}, R., {Schwope}, A.~D., {Vogel}, J., {et~al.} 2009, \aap, 496, 833

\bibitem[{{Schwope} {et~al.}(2002){Schwope}, {Brunner}, {Buckley}, {Greiner},
  {Heyden}, {Neizvestny}, {Potter}, \& {Schwarz}}]{schwope:02}
{Schwope}, A.~D., {Brunner}, H., {Buckley}, D., {et~al.} 2002, \aap, 396, 895

\bibitem[{{Schwope} {et~al.}(2003){Schwope}, {Thomas}, {Mante}, {Haefner}, \&
  {Staude}}]{schwope:03}
{Schwope}, A.~D., {Thomas}, H.-C., {Mante}, K.-H., {Haefner}, R., \& {Staude},
  A. 2003, \aap, 402, 201

\bibitem[{{Silva} {et~al.}(2011){Silva}, {Rodrigues}, \& {Costa}}]{silva:12pre}
{Silva}, K.~M.~G., {Rodrigues}, C.~V., \& {Costa}, J.~E.~R. 2011,
  [arXiv:astro-ph/1101.5568]

\bibitem[{{Sion}(1999)}]{sion:99}
{Sion}, E.~M. 1999, \pasp, 111, 532

\bibitem[{{Sirk} \& {Howell}(1998)}]{sirk:98}
{Sirk}, M.~M. \& {Howell}, S.~B. 1998, \apj, 506, 824

\bibitem[{{Standish}(1998)}]{standish:98}
{Standish}, E.~M. 1998, JPL IOM, 312.F-98-048

\bibitem[{{Staude} {et~al.}(2001){Staude}, {Schwope}, \& {Schwarz}}]{staude:01}
{Staude}, A., {Schwope}, A.~D., \& {Schwarz}, R. 2001, \aap, 374, 588

\bibitem[{{Szkody} {et~al.}(1999){Szkody}, {Vennes}, {Schmidt}, {Wagner},
  {Fried}, {Shafter}, \& {Fierce}}]{szkody:99}
{Szkody}, P., {Vennes}, S., {Schmidt}, G.~D., {et~al.} 1999, \apj, 520, 841

\bibitem[{{Thomas} {et~al.}(1998){Thomas}, {Beuermann}, {Reinsch}, {Schwope},
  {Truemper}, \& {Voges}}]{thomas:98}
{Thomas}, H.-C., {Beuermann}, K., {Reinsch}, K., {et~al.} 1998, \aap, 335, 467

\bibitem[{{Townsley} \& {G{\"a}nsicke}(2009)}]{townsley:09}
{Townsley}, D.~M. \& {G{\"a}nsicke}, B.~T. 2009, \apj, 693, 1007

\bibitem[{{Traulsen} {et~al.}(2010){Traulsen}, {Reinsch}, {Schwarz},
  {Dreizler}, {Beuermann}, {Schwope}, \& {Burwitz}}]{traulsen:10}
{Traulsen}, I., {Reinsch}, K., {Schwarz}, R., {et~al.} 2010, A\&A, 516, A76

\bibitem[{{Traulsen} {et~al.}(2011){Traulsen}, {Reinsch}, {Schwope}, {Burwitz},
  {Dreizler}, {Schwarz}, \& {Walter}}]{traulsen:11}
{Traulsen}, I., {Reinsch}, K., {Schwope}, A.~D., {et~al.} 2011, \aap, 529, A116

\bibitem[{{van Teeseling} {et~al.}(1996){van Teeseling}, {Kaastra}, \&
  {Heise}}]{teeseling:96}
{van Teeseling}, A., {Kaastra}, J.~S., \& {Heise}, J. 1996, \aap, 312, 186

\bibitem[{{Vennes} {et~al.}(1995){Vennes}, {Szkody}, {Sion}, \&
  {Long}}]{vennes:95}
{Vennes}, S., {Szkody}, P., {Sion}, E.~M., \& {Long}, K.~S. 1995, \apj, 445,
  921

\bibitem[{{Verner} \& {Ferland}(1996)}]{verner:96a}
{Verner}, D.~A. \& {Ferland}, G.~J. 1996, \apjs, 103, 467

\bibitem[{{Verner} {et~al.}(1996){Verner}, {Ferland}, {Korista}, \&
  {Yakovlev}}]{verner:96b}
{Verner}, D.~A., {Ferland}, G.~J., {Korista}, K.~T., \& {Yakovlev}, D.~G. 1996,
  \apj, 465, 487

\bibitem[{{Voges} {et~al.}(1999){Voges}, {Aschenbach}, {Boller},
  {Br{\"a}uninger}, {Briel}, {Burkert}, {Dennerl}, {Englhauser}, {Gruber},
  {Haberl}, {Hartner}, {Hasinger}, {K{\"u}rster}, {Pfeffermann}, {Pietsch},
  {Predehl}, {Rosso}, {Schmitt}, {Tr{\"u}mper}, \& {Zimmermann}}]{voges:99}
{Voges}, W., {Aschenbach}, B., {Boller}, T., {et~al.} 1999, \aap, 349, 389

\bibitem[{{Werner}(1986)}]{werner:86}
{Werner}, K. 1986, \aap, 161, 177

\bibitem[{{Werner} \& {Dreizler}(1999)}]{werner:99}
{Werner}, K. \& {Dreizler}, S. 1999, J.\ Comp.\ Appl.\ Math., 109, 65

\bibitem[{{Wilms} {et~al.}(2000){Wilms}, {Allen}, \& {McCray}}]{wilms:00}
{Wilms}, J., {Allen}, A., \& {McCray}, R. 2000, \apj, 542, 914

\end{thebibliography}

\end{document}